\documentclass[pdflatex,sn-mathphys]{sn-jnl}

\jyear{2021}%

\usepackage{multicol}
\usepackage{stackengine}
\newcommand{\Dbar}{\stackinset{l}{0.1ex}{c}{}{\rule{0.33em}{0.3pt}}{D}}

\theoremstyle{thmstyleone}%
\theoremstyle{thmstyletwo}%

\theoremstyle{thmstylethree}%

\begin{document}

\title[ ]{Rheological properties of acid-induced carboxymethylcellulose hydrogels}


\author[1]{\fnm{Gauthier} \sur{Legrand}}

\author[2]{\fnm{Guilhem P.} \sur{Baeza}}

\author[1,3]{\fnm{S\'ebastien} \sur{Manneville}}%

\author*[1]{\fnm{Thibaut} \sur{Divoux}}
\email{Thibaut.Divoux@ens-lyon.fr}

\affil[1]{ENSL, CNRS, Laboratoire de physique, F-69342 Lyon, France}

\affil[2]{Univ Lyon, INSA Lyon, UCBL, CNRS, MATEIS, UMR5510, 69621, Villeurbanne, France}

\affil[3]{Institut Universitaire de France (IUF)}


\abstract{Cellulose ethers represent a class of water-soluble polymers widely utilized across diverse sectors, spanning from healthcare to the construction industry. This experimental study specifically delves into aqueous suspensions of carboxymethylcellulose (CMC), a polymer that undergoes gel formation in acidic environments due to attractive interactions between hydrophobic patches along its molecular chain. We use rheometry to determine the linear viscoelastic properties of both CMC suspensions and acid-induced gels at various temperatures. Then, applying the time-temperature superposition principle, we construct master curves for the viscoelastic spectra, effectively described by fractional models. The horizontal shift factors exhibit an Arrhenius-like temperature dependence, allowing us to extract activation energies compatible with hydrophobic interactions. Furthermore, we show that acid-induced CMC gels are physical gels that display a reversible yielding transition under external shear. In particular, we discuss the influence of pH on the non-linear viscoelastic response under large-amplitude oscillatory shear. Overall, our results offer a comprehensive description of the linear and non-linear rheological properties of a compelling case of physical hydrogel involving hydrophobic interactions. 
}


\maketitle

\section{Introduction}
\label{sec:Introduction}

Physical gels are ubiquitously found in a broad range of products, from foodstuffs to biomedical products. They consist of
assemblies of colloids or polymers displaying attractive interactions, hence forming a percolated network incorporating large amounts of solvent \cite{Zhang:2017,Cao:2020}. The very nature of the attractive forces between the constituents plays a key role in the gel properties, for they impact the strength and connectivity of the gel network. In practice, attractive interactions come in different flavors usually sorted by their strength \cite{Rossow:2015}. Physical gels built from polymers involve various types of non-covalent bonds, including metal-ligand \cite{Khare:2021, jiang2022magneto, zhuge2017decoding}, hydrogen bonds \cite{cui2018linear,louhichi2017humidity,baeza2016network,bouteiller2007assembly}, and ionic bonds \cite{shabbir2017nonlinear, chen2013ionomer, mei2022anion}, as well other types of attractive interactions, such as van der Waals interactions and so-called hydrophobic interactions \cite{Tuncaboylu:2012}. The latter case is of primary importance since hydrophobic interactions are pivotal in the self-assembly of numerous biological systems, e.g., the structure of proteins and DNA \cite{Cerny:2007,Wang:2017}. However, gels formed by hydrophobic interactions remain less studied than their counterpart formed from the other interactions listed above, and their specific characteristics are yet to be fully understood.   

Here we tackle the case of hydrogels formed by hydrophobic interactions between carboxymethylcellulose (CMC), a polyelectrolyte also known as cellulose gum and commonly encountered {as} a food additive, registered as E466 or E469. In practice, CMC is obtained from natural cellulose --a polymer that is not hydrosoluble--  by etherification, using sodium hydroxide and monochloroacetate \cite{Heinze:1999}. As a result, each anhydroglucose repeating unit of CMC displays up to three hydroxyl groups substituted by a carboxymethyl group. Such substitution is scattered along the polymer chain, yielding hydrophobic regions, which correspond to the less substituted ones \cite{Debutts:1957}. The \textit{average} number of substituted hydroxyl groups per anhydroglucose repeating unit defines the degree of substitution (DS) of the CMC, which typically ranges between 0.5 and 1.5 for commercial grades. Blocks of unsubstituted cellulose regions along the chain backbone are mainly observed for DS~$\lesssim 0.9$ \cite{Lopez:2018}. These unsubstituted regions facilitate transient chain associations, responsible for a pronounced thixotropic behavior under flow, and the formation of gels at large concentrations \cite{Debutts:1957,Elliot:1974,Barba:2002,Lopez:2018,Lopez:2021}.

At fixed CMC content, gel formation in suspensions of weakly substituted CMC can further be controlled by adjusting the pH of the suspension. Indeed, lowering the pH of the CMC suspension diminishes the charge density of the polymer chains, thereby favoring polymer-polymer interactions and promoting the formation of multi-chain aggregates \cite{Dogsa:2014}. In a recent study, we have provided a comprehensive description of such an acid-induced gelation of carboxymethylcellulose \cite{Legrand:2024}. In particular, we reported a phase diagram in the (pH-CMC concentration) plane, which shows a gel region at low pH for suspensions exceeding 1\%~wt. Building upon seminal observations \cite{Durig:1950,Hermans:1965},  we introduced a mean-field model based on hydrophobic interchain association that accounts remarkably well for the sol-gel boundary over a broad range of CMC concentrations, up to 5\% (Fig.~\ref{fig:PD}). Finally, our neutron scattering experiments suggest the picture of a gel network built from flexible rod-like structures \cite{Legrand:2024}. In this network, the effective crosslinks consist of fuzzy colloidal structures composed of a dense core of aggregated polymers surrounded by a sparse, hairy shell of polymer chains \cite{Legrand:2024}. Such crosslinkers are strongly reminiscent of fringed micelles, previously identified by atomic force microscopy \cite{Liebert:2001,Liebert:2005}.

In the present article, we focus on the rheological properties of sodium carboxymethylcellulose (NaCMC) suspensions across the sol-gel transition. After presenting the materials and methods in section~\ref{sec:Materials}, we show in section~\ref{sec:Results} that the linear viscoelastic spectrum of NaCMC suspensions in the liquid phase displays a broad power-law behavior, which can be rescaled onto a master curve following a time-temperature superposition principle. The horizontal shift factor obeys an Arrhenius dependence with the temperature, which allows us to infer activation energy that is compatible with hydrophobic interactions. A similar analysis in the gel phase reveals that acid-induced NaCMC gels follow the theory of rubber elasticity. Finally, in section~\ref{sec:nonlinear} of the manuscript, we examine the shear-induced yielding transition of NaCMC gels, which display a response prototypical of yield stress fluids \cite{Bonn:2017}. In particular, Large Amplitude Oscillatory Shear reveals that NaCMC gels exhibit enhanced dissipation at the yield point, which hints at a spatially heterogeneous solid-to-liquid transition. Further intra-cycle analysis offers a picture at the microscale of such a yielding transition that we discuss at various pH. Our results offer a comprehensive description of the linear and non-linear rheological properties of carboxymethylcellulose physical hydrogels built from enhanced hydrophobic interactions at low pH. 

\section{Materials and methods}
\label{sec:Materials}

\subsection{Sample preparation}

CMC suspensions are prepared by dissolving a sodium salt of carboxymethylcellulose (NaCMC, Sigma Aldrich, $M_w=250~\rm kg.mol^{-1}$ and $\mathrm{DS}=0.9$ as specified by the manufacturer) in deionized water. Actual values for the batch under study were determined to be $M_w=213~\rm kg.mol^{-1}$ with a polymolecularity index $\Dbar=2.55$, and $\mathrm{DS}=0.88$ using size exclusion chromatography and high-field NMR spectroscopy, respectively \cite{Legrand:2024}. Furthermore, we have verified by determining the specific viscosity of NaCMC suspensions over a broad range of concentrations that the transition from the semi-dilute unentangled to entangled occurs at $c_e=0.24\%$, while the transition from the semi-dilute entangled and the concentrated regime occurs at $c^{**}=2.4\%$ \cite{Lopez:2018}.
Stock solutions up to 5\% wt.~are prepared and stirred at room temperature for 48~hours until homogeneous before diluting them with a hydrochloric acid solution (Sigma Aldrich) and mechanically agitated for 7~days on a bottle roller. Final CMC solutions concentrations span from 1~\% to 4.5~\%. We used a 1~M HCl solution for all the dilutions except for the most concentrated CMC solutions (4 and 4.5~\%) for which we used a 12~M HCl solution. The final pH of the samples, measured using a pH-meter (Mettler Toledo SevenCompact) calibrated with three buffer solutions (pH=2, 4, and 7 from Mettler Toledo) spans between 0 and 7. 

\subsection{Rheometry}
\label{sec:rheometry}
The rheological properties are determined with a cone-and-plate geometry (angle 2$^\circ$, radius $20~\rm mm$ and truncation $46~\rm \mu$m) connected to a strain-controlled rheometer (ARES G2, TA Instruments). The cone and plate are sandblasted and display a surface roughness of about 1~$\mu$m to prevent wall slip. Samples are loaded in the shear cell, and maintained at constant temperature $T$ for the whole duration of the test by a Peltier modulus placed under the bottom plate. The rheological protocol applied to the samples for linear viscoelastic characterization is divided into three consecutive steps: ($i$) a preshear at $\dot \gamma=50$~s$^{-1}$ for $3~\rm min$ to erase the loading history and rejuvenate the sample \cite{Viasnoff:2002,Bonn:2017,Joshi:2018}; ($ii$) a recovery phase of $20~\rm min$ during which we monitor the sample linear viscoelastic properties by applying small amplitude oscillations with amplitude $\gamma_0=1$\% and frequency $\omega = 2\pi ~\rm rad.s^{-1}$; note that a steady state is reached typically within $100~\rm s$; ($iii$) a frequency sweep performed by ramping the frequency from $\omega = 2 \pi.10^2 ~\rm rad.s^{-1}$ down to $\omega = 2 \pi.10^{-2}~\rm rad.s^{-1}$ in logarithmic steps at $\gamma_0=1$\% to determine the linear viscoelastic spectrum of the CMC suspension over three decades in frequency.
In the following, $T$ is chosen between $5^\circ$C to $40^\circ$C in order to test for the influence of temperature on the sample viscoelastic properties. The protocols used for characterizing the samples in the nonlinear regime will be specified in the dedicated section below.

\section{Linear viscoelasticity of CMC suspensions}
\label{sec:Results}

\begin{figure}[!t]
\centering
\includegraphics[width = 0.5\linewidth]{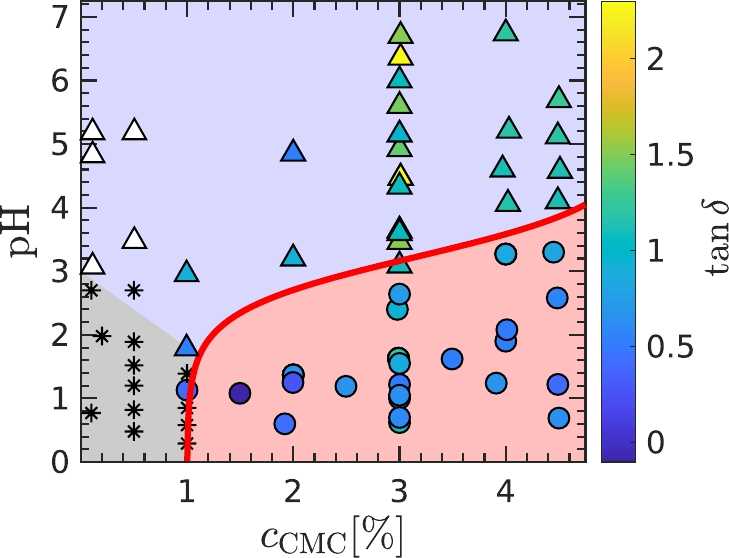}
\caption{Phase diagram for aqueous CMC suspensions at $T=22^\circ \rm C$ in the pH vs. CMC weight fraction plane. Symbols denote four different phases: gel [($\bullet$) with a red background], viscoelastic liquid [($\Delta$) with a blue background], and phase-separated samples [($\star$) with a gray background]. Color levels code for the loss factor $\tan \delta=G''/G'$ of the gel sample determined by small amplitude oscillatory shear at $\omega=2 \pi$~rad.s$^{-1}$. The white triangles denote viscoelastic homogeneous liquids whose rheological properties were not measured. The red curve separating the viscoelastic liquid phase from the gel phase corresponds to the model discussed in detail in ref.~\cite{Legrand:2024}.}\label{fig:PD}
\end{figure}

\subsection{Phase diagram}

Aqueous suspensions of carboxymethylcellulose form gels upon acidification. In a recent study, we have performed an extensive characterization of the sol-gel transition varying both the CMC content and the pH of these solutions \cite{Legrand:2024}. Measuring the linear viscoelastic spectrum of these suspensions, i.e., $G'(\omega)$ and $G''(\omega)$ allows us to identify the gel point, which corresponds to sample compositions such that $\tan \delta = G''/G'$ is frequency independent \cite{Winter:1997}, while gel samples are defined by $\tan \delta <1$ in the limit $\omega \rightarrow 0$. This criterion allowed us to build the phase diagram reported for $T=22^\circ \rm C$ in Fig.~\ref{fig:PD} in which the gel phase corresponds to CMC suspensions of concentration larger than 1\% and sufficiently low pH to induce the aggregation of the CMC. Suspensions with a CMC concentration lower than 1\% do not form gels but turbid heterogeneous dispersions of macroscopic aggregates due to the lack of polymer (see gray region in Fig.~\ref{fig:PD}), whereas CMC suspensions of sufficiently high pH form homogeneous viscoelastic liquids (see blue region in Fig.~\ref{fig:PD}). In the following, we examine in more detail the linear viscoelastic spectrum of both the liquid and gel phases and test their robustness to a change in temperature.

\subsection{Time-temperature superposition}
 
\subsubsection{Analysis in the sol phase}
\begin{figure*}[!t]
\centering
\includegraphics[width = 0.9\linewidth]{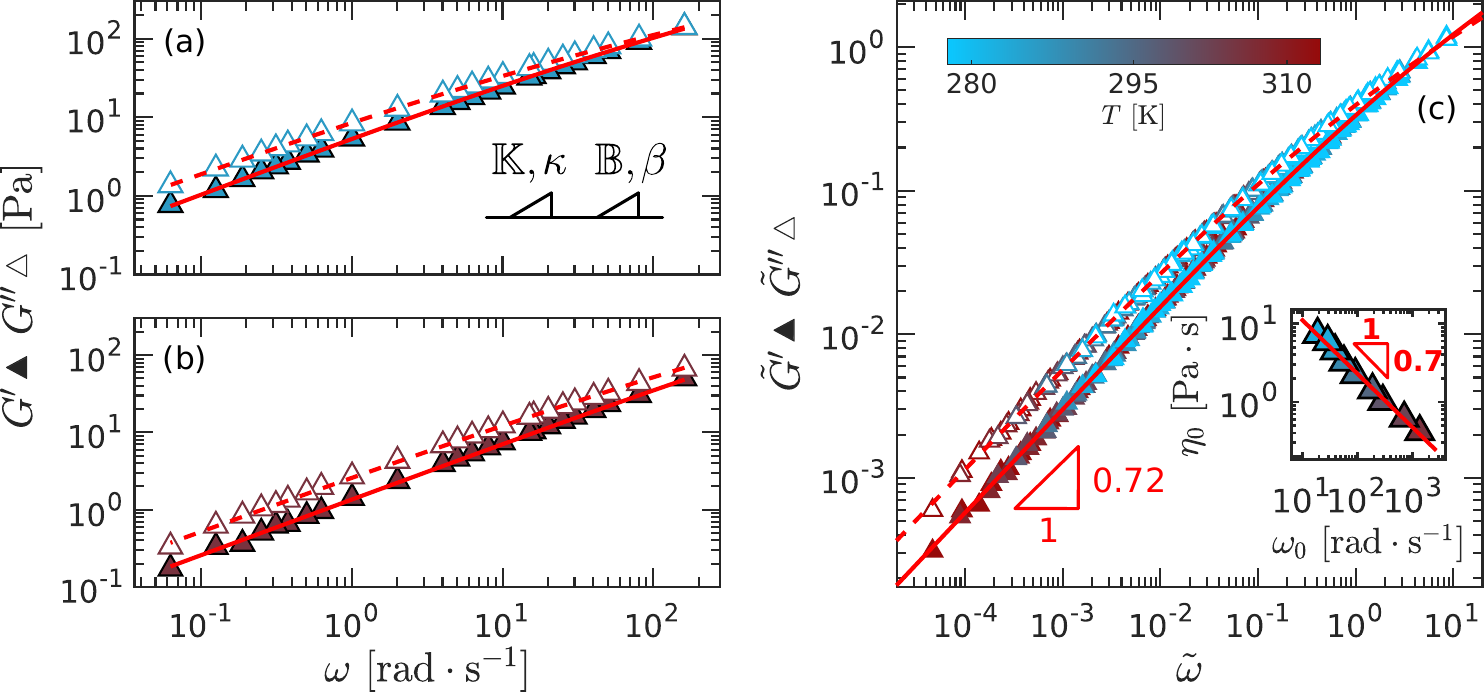}
\caption{
Viscoelastic spectrum in the liquid phase: frequency dependence of the elastic and viscous moduli, $G'$ ($\blacktriangle$) and $G''$ ($\triangle$) resp., of a CMC solution with $c_{\rm CMC}=3\%$ and $\textrm{pH}=3.6$ (a) at $T=5^\circ \rm C$ and (b) at $T=40^\circ \rm C$. The red curves are the best fits of the data to a fractional Maxwell (FM) model [see Eq.~\eqref{eq:FM} and sketch in (a)].
(c) Master curve for the frequency dependence of the viscoelastic moduli obtained by normalizing both the moduli and the frequency: $\tilde{G}'=G'/(\eta_0 \omega_0)$, $\tilde{G}''=G''/(\eta_0 \omega_0)$, and $\tilde{ \omega} = \omega/\omega_0$ with 
$\eta_0 = ( \mathbb{K}^{\beta -1} / \mathbb{B}^{\kappa - 1})^{1 / (\beta-\kappa)} $ and $\omega_0 = ( \mathbb{K} / \mathbb{B})^{1 / (\beta - \kappa)}$ for eight different temperatures ranging from $T=5^\circ \rm C$ to 40$^\circ \rm C$ as indicated by the color bar.  The red curves correspond to the normalized FM model: $\tilde{G}^\star = \left( i \tilde{\omega} \right)^{\kappa + \beta} / [ \left( i \tilde{\omega} \right)^{\kappa} + \left( i \tilde{\omega} \right)^{\beta} ]$ with $\kappa = 0.39$ and $\beta = 0.72$. Inset: $\eta_0$ vs.~$\omega_0$; the red line is the best power-law fit of the data, $\eta_0\sim\omega_0^{-0.7}$.  
}\label{fig:FMM}
\end{figure*}

We first examine the linear viscoelastic properties of acidified CMC suspensions in the liquid phase, and the concentrated regime. Figure~\ref{fig:FMM}(a) illustrates the frequency dependence of $G'$ and $G''$ in the case of a liquid 3\% CMC suspension at $\rm pH=3.6$ and $T=5^\circ$C. The viscoelastic spectrum shows a broad power-law behavior, in line with previous rheological measurements performed on CMC suspensions at similar concentrations for various counter ions and for CMC of different molecular properties, i.e., various degrees of substitution and molecular weights \cite{Matsumoto:1988,Kastner:1997,Barba:2002,Benchabane:2008,Lopez:2019,Lopez:2021}. The power-law viscoelastic response of the CMC suspension is remarkably well-captured by a fractional Maxwell (FM) model that consists of two spring pots -- or Scott-Blair elements-- in series [see sketch as an inset in Fig.~\ref{fig:FMM}(a)]. Each spring pot is characterized by a quasi-property and a dimensionless exponent, noted ($\mathbb{K}$, $\kappa$) and ($\mathbb{B}$, $\beta$), respectively, where we assume $\kappa<\beta$. The corresponding complex modulus reads \cite{Jaishankar:2013,Bonfanti:2020}:
\begin{equation}
G^\star(\omega) = \dfrac{ \mathbb{K}\left( i \omega \right)^{\kappa } \mathbb{B}\left( i \omega \right)^{\beta}}{ \mathbb{K} \left( i \omega \right)^{\kappa} + \mathbb{B}\left( i \omega \right)^{\beta} }
    \label{eq:FM}
\end{equation}
The complex modulus can be re-written as follows:
\begin{equation}
G^\star(\omega) = \eta_0 \omega \dfrac{\left( i \omega / \omega_0 \right)^{\kappa +\beta-1}}{ \left( i \omega / \omega_0\right)^{\kappa} + \left( i \omega / \omega_0 \right)^{\beta} }
    \label{eq:FM_norm}
\end{equation}
where $\eta_0 = ( \mathbb{K}^{\beta -1} / \mathbb{B}^{\kappa - 1})^{1 / (\beta-\kappa)} $ and $\omega_0 = ( \mathbb{K} / \mathbb{B})^{1 / (\beta - \kappa)}$. The best fits of the data with a FM model yield $\beta=0.72\pm0.02$ and $\kappa=0.39\pm0.04$. Therefore, in the low-frequency limit, i.e., $\omega \ll \omega_0$, the CMC suspension displays a critical-like response, $G'(\omega)\sim G''(\omega)\sim \omega^{0.7}$. This scaling strongly contrasts with the response expected for a monodisperse polymer solution, for which $G''(\omega)\sim \omega$, while $G'(\omega)\sim \omega^2$ \cite{Larson:1999}. Such a discrepancy can be related to the presence of hydrophobic patches along the polymer chains that lead to some degree of physical binding between the polymers, bringing the suspension close to the gel point.

This response is robustly observed for temperatures ranging from $T=5^\circ$C to $40^\circ$C [see Fig.~\ref{fig:FMM}(b)], which hints at some universal feature in the linear mechanical response of CMC suspensions. Indeed, the viscoelastic spectrum obtained at various temperatures can be rescaled into a master curve by normalizing the elastic and viscous moduli by $\eta_0\omega_0$, and the frequency by $\omega_0$, as illustrated in Fig.~\ref{fig:FMM}(c). Such a rescaling highlights the existence of a time-temperature superposition, which is here in the terminal regime governed by the dynamical association between CMC strands as originally proposed in the sticky reptation model \cite{leibler1991dynamics,rubinstein2001dynamics}. Moreover, we find that these two rescaling factors are connected by a power law that reads $\eta_0 \sim \omega_0^{-0.7}$ or equivalently, in terms of elastic scale, $\eta_0 \omega_0 \sim \omega_0^{0.3}$. In that framework, the parameters $\omega_0^{-1}$ and $(\eta_0 \omega_0)^{-1}$ are equivalent to the horizontal and vertical shift factors, respectively $a_T$ and $b_T$ usually employed for time-temperature superposition \cite{rubinstein2003polymer}. 

\begin{figure*}[!t]
\centering
\includegraphics[width = 0.9\linewidth]{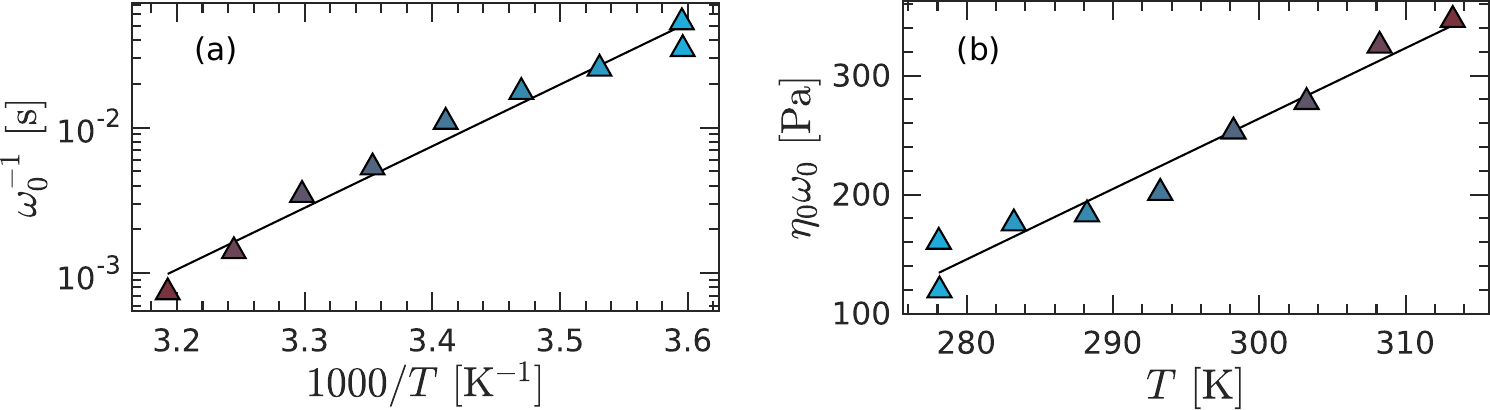}
\caption{Temperature dependence of the scaling factors $\omega_0$ and $\eta_0$ extracted from the fractional Maxwell model of a CMC solution with $c_{\rm CMC}=3\%$ and $\mathrm{pH}=3.6$ shown in Fig.~\ref{fig:FMM}. (a) $\omega_0^{-1}$ vs.~$1/T$ plotted in semi-logarithmic scales to highlight the exponential scaling.  The black line corresponds to the best fit of the data with an Arrhenius-like scaling, $\omega_0^{-1}\sim \exp(-E_a/k_BT)$, where $E_a$ stands for the activation energy, and $k_B$ is the Boltzmann constant. The best fit is obtained with $E_a=81\pm 9~\rm kJ/mol$. (b) $(\eta_0 \omega_0)$ vs.~$T$. The black line stands for the best linear fit of the data. Colors code for the temperature. Same color code as in Fig.~\ref{fig:FMM}.}\label{fig:Arrhenius_FMM}
\end{figure*}

Interestingly, similar master curves were reported for CMC suspensions prepared with various salt concentrations, and various types of monovalent salts, e.g., NaCl, and divalent salts, e.g., CaCl$_2$, MgCl$_2$, etc., which hints at a time-salt superposition principle \cite{Matsumoto:1988}. Yet, these master curves were obtained by shifting manually the viscoelastic spectra along the sole horizontal --frequency-- axis, with a shift factor that depends on the nature of the salt. 
The microscopic scenario underpinning this \textit{simple} superposition principle has been attributed to an increase of ``segmental friction" of the polymer chain due to the ion complexation with the carboxyl group, at least for calcium ions \cite{Matsumoto:1990,Matsumoto:1992}.


Here, superposition is obtained by shifting both horizontally and vertically the viscoelastic spectrum, which suggests that varying the temperature affects not only the solution dynamics but also the number of physical crosslinks between the CMC chains. Indeed, the elasticity scale $\eta_0\omega_0$ is an increasing function of temperature. More precisely,  the vertical shift factor $(\eta_0\omega_0)^{-1}$ is found to decrease by a factor \textit{ca.} 2.7 when increasing the temperature from $5 \rm ^{\circ}C$ to $40 \rm ^{\circ}C$ [see Fig. \ref{fig:Arrhenius_FMM}(a)], which must be related to a weak, albeit significant, additional association of the CMC hydrophobic patches. This corresponds to a slightly lower solvent quality upon increasing the temperature and is also detectable in the gel phase, beyond entropic elasticity considerations (see below). Conversely, the horizontal shift factor $\omega_0^{-1}$ [see Fig. \ref{fig:Arrhenius_FMM}(b)], which corresponds to the inverse of the terminal relaxation time, is well described by an Arrhenius equation providing an activation energy of \textit{ca.} $81\ \rm kJ/mol$. This value is significantly higher than what is generally observed in polymer melts far above their glass transition temperature ($\approx\ 50~\rm kJ/mol$) \cite{rubinstein2003polymer}. Such a result is compatible with the presence of additional associative bonds between the CMC chains.

\begin{figure*}[!t]
\centering
\includegraphics[width = 0.9\linewidth]{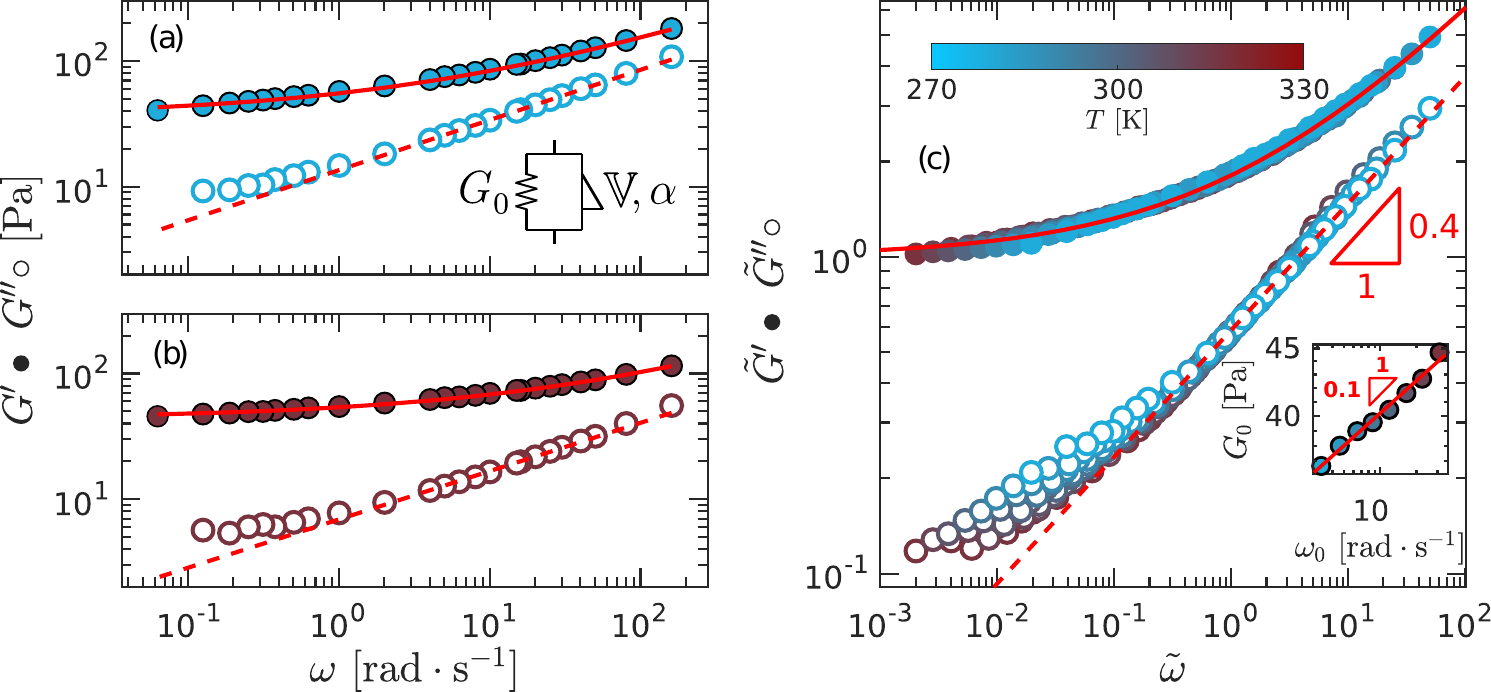}
\caption{Viscoelastic spectrum in the gel phase: frequency dependence of the elastic and viscous moduli, $G'$ ($\bullet$) and $G''$ ($\circ$) resp., of a CMC gel with $c_{\rm CMC}=3\%$ and $\mathrm{pH}=1.6$ at (a) $T=5^\circ \rm C$ and (b) at $T=40^\circ \rm C$. The red curves are the best fits of the data to a fractional Kelvin-Voigt (FKV) model [see Eq.~\eqref{eq:FKV} in the main text and sketch in (a)]. (c) Master curve for the frequency dependence of the viscoelastic moduli obtained by normalizing both the viscoelastic moduli and the frequency, i.e., $\tilde{G}'=G'/G_0$, $\tilde{G}''=G''/G_0$, and $\tilde{ \omega} = \omega/\omega_0$ with $\omega_0=(G_0/\mathbb{V})^{1/\alpha}$ for eight different temperatures ranging from $T=5^\circ \rm C$ to $40^\circ \rm C$ as indicated by the color bar. The red curves correspond to the normalized FKV model, $\tilde{G}^\star = 1 + \left( i \tilde{\omega} \right)^\alpha$ with $\alpha = 0.40\pm 0.01$. Inset: $G_0$ vs.~$\omega_0$; the red line is the best power-law fit of the data, $G_0\sim \omega_0^{0.1}$.}\label{fig:FKV}
\end{figure*}

\subsubsection{Analysis in the gel phase}

We now turn to the linear viscoelastic properties of acidified CMC suspensions in the gel phase. The viscoelastic spectrum of a 3\% CMC gel at $\rm pH=1.6$ is shown in Fig.~\ref{fig:FKV}(a). The elastic modulus is larger than the viscous modulus, including in the limit of vanishing frequencies, where $G'$ converges to a plateau value. This confirms the solid-like behavior of the 3\% CMC suspension at this pH. Moreover, both $G'$ and $G''$ display a power-law-like response at high frequencies. Such a frequency dependence is well-captured by a fractional Kelvin-Voigt (FKV) model, which consists of a spring of shear modulus $G_0$ in parallel with a single spring pot characterized by a quasi-property $\mathbb{V}$ and a dimensionless exponent $\alpha$ [see sketch in Fig.~\ref{fig:FKV}(a)]. The complex modulus of the FKV model reads:
\begin{equation}
    G^\star(\omega) = G_0 + \mathbb{V}(i\omega)^\alpha
    \label{eq:FKV}
\end{equation}
from which we use the real and imaginary parts to jointly fit the viscoelastic spectrum. Except for $G''$ at the lowest frequencies, the model shows an excellent agreement with the data at various temperatures [see Fig.~\ref{fig:FKV}(a) and \ref{fig:FKV}(b)]. Interestingly, the power-law exponent $\alpha$ is independent of the temperature, which prompts us to set it constant $\alpha=0.4$, which reduces the number of fitting parameters to two: $G_0$ and $\mathbb{V}$. The latter parameter can thus be replaced by $\omega_0=(G_0/\mathbb{V})^{1/\alpha}$, which has a physical dimension much easier to grasp than $\mathbb{V}$. Moreover, using normalized coordinates, i.e., $\tilde G'=G'/G_0$, $\tilde G''=G''/G_0$, and $\tilde \omega =\omega/\omega_0$, we obtain a master curve for the viscoelastic spectra measured at various temperatures. Again, we emphasize that the traditional horizontal and vertical shift factors in time-temperature superposition correspond, respectively, to $a_T=\omega_0^{-1}$ and $b_T=G_0^{-1}$.

\begin{figure*}[!t]
\centering
\includegraphics[width = 0.9\linewidth]{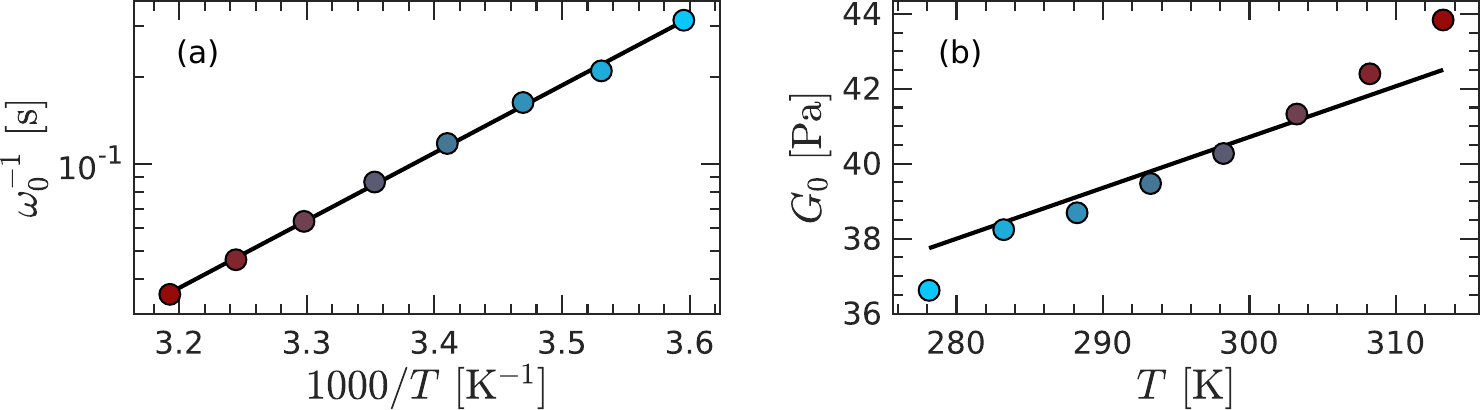}
\caption{Temperature dependence of the scaling factors $G_0$ and $\omega_0$ from the fractional Kelvin-Voigt model of a CMC gel with $c_{\rm CMC}=3\%$ and $\mathrm{pH}=1.6$ shown in Fig.~\ref{fig:FKV}. (a) $\omega_0^{-1}$ vs.~$1/T$. The black line is the best fit of the data with an Arrhenius-like scaling, $\omega_0^{-1}\propto \exp (-E_a/k_BT)$, yielding an activation energy $E_a=43\pm 3~\rm kJ/mol$. (b) $G_0$ vs. $T$. The black line shows the best fit of the data assuming a rubber elasticity model, i.e., $G_0 = k_B T / \xi^3$, yielding $\xi \simeq 46~\rm nm$.  Colors code for the temperature. Same color code as in Fig.~\ref{fig:FKV}. }\label{fig:param_KVF_vs_T}
\end{figure*}

To the best of our knowledge, this is the first time that such time-temperature superposition is reported for acid-induced CMC gels. More quantitatively, the high-frequency exponent $\alpha=0.40$ is strongly reminiscent of the power-law exponent $\kappa=0.39$ describing the high-frequency linear viscoelastic behavior in the liquid state. This observation suggests that $\alpha$ and $\kappa$ could be a single exponent constant across the sol-gel transition. In general, the high-frequency response pertains to small length scales where both liquid CMC suspensions and CMC gels should be structurally similar, i.e., composed of patchy polymers experiencing some degree of attractive interactions. 

Furthermore, in previous work, we obtained a similar master curve by varying the pH at fixed temperature and CMC concentration \cite{Legrand:2024}. This suggests that pH and temperature affect the same binding element, which are the hydrophobic groups along the polymer chain. Note that the specific values of the exponent, here $\alpha=0.40\pm 0.01$ vs. $\alpha=1/2$ in Ref.~\cite{Legrand:2024}, differ significantly. As shown in Fig.~S1 in the Supplementary Information, this discrepancy can be ascribed to a slow reinforcement of acid-induced CMC gels over the course of several months, which results in a slow decrease of $\alpha$ from $\alpha=0.55$ just a few days after sample preparation to $\alpha =0.40$ two months later.

Finally, we examine the temperature dependence of the rescaling factors, i.e., $\omega_0$ and $G_0$. 
On the one hand, the horizontal rescaling factor $\omega_0^{-1}$ obeys an Arrhenius dependence governed by the interactions between the polymer and the solvent. Yet, surprisingly, the temperature dependence yields an activation energy $E_a \simeq 43~\rm kJ/mol$ that is \textit{ca.} twice as small as the energy computed from the CMC suspensions in the liquid state (see Fig.~\ref{fig:Arrhenius_FMM}). Such a discrepancy undoubtedly originates from a change in the dominant relaxation mechanism governing the viscoelastic properties in the gel state. While in the liquid state, the dynamics are mainly governed by the association/dissociation of the hydrophobic patches, the larger size and concentration of fringe micelles at low pH results in arrested dynamics at low frequencies, i.e., in the formation of a network. In addition, it is interesting to remark that additional relaxation modes emerge at high frequencies with a characteristic dependence of $\Tilde{G'}\sim \Tilde{G''}\sim \Tilde{\omega}^{0.4}$, reminiscent of a Rouse-like relaxation yet with a power-law exponent smaller than $0.5$. In other words, the viscoelastic spectrum of the gel can be separated into two domains: for $\Tilde{\omega}<10^{-1}$, the network made of fringe micelles connected by CMC strands dominates the rheological response, whereas for $\Tilde{\omega}>10^{-1}$, the response originates from the Rouse-like relaxation of CMC strands, being controlled by the polymer-solvent interactions at a smaller length scale. This interpretation is further supported by the partial failure of the time-temperature superposition principle at intermediate frequencies [see particularly $\Tilde{G''}$ for $\Tilde{\omega}<10^{-1}$ in Fig.~\ref{fig:FKV}(c)], which is typical of thermorheologically complex systems such as supramolecular polymer networks including double networks \cite{cui2018linear, stadler2009linear, coutouly2024low, lyons2022equilibration}. From a more quantitative point of view, the value of the activation energy used to form the gel master curve at high frequency ($E_a\simeq 43\ \rm kJ/mol$) is in good agreement with what is expected for polymer melts far above their glass transition temperature ($\approx 50~\rm kJ/mol$) \cite{rubinstein2003polymer}.

On the other hand, the elastic modulus $G_0$ increases with the temperature in a way that is compatible with the theory of rubber elasticity, i.e., $G_0 = k_BT/\xi^3$, where $\xi$ denotes the average distance between two nodes of the gel network [Fig.~\ref{fig:param_KVF_vs_T}(a)]. Here, the best fit of the data yields $\xi \simeq 46~\rm nm$. This length scale is comparable with the size of fringed micelles, as determined recently by neutron scattering on the very same samples \cite{Legrand:2024}. 
This observation suggests that fringed micelles act as crosslinkers within the CMC hydrogel. 
Note that seminal work on CMC hydrogels with $0.8\lesssim\mathrm{DS}\lesssim 1$, and formed at sufficiently large CMC concentrations have shown that the dependence of the elastic modulus with the polymer concentration is compatible with the theory of rubber elasticity \cite{Hermans:1965}. Our results, therefore, extend these conclusions to the case of acid-induced CMC gels (at fixed CMC concentration), while explicitly testing the temperature dependence, which, to the best of our knowledge, had not been reported yet.

\section{Non-linear rheological properties of acid-induced CMC gels}
\label{sec:nonlinear}

\subsection{Reversibility of the yielding transition}

\begin{figure*}[!t]
\centering
\includegraphics[width = 0.9\linewidth]{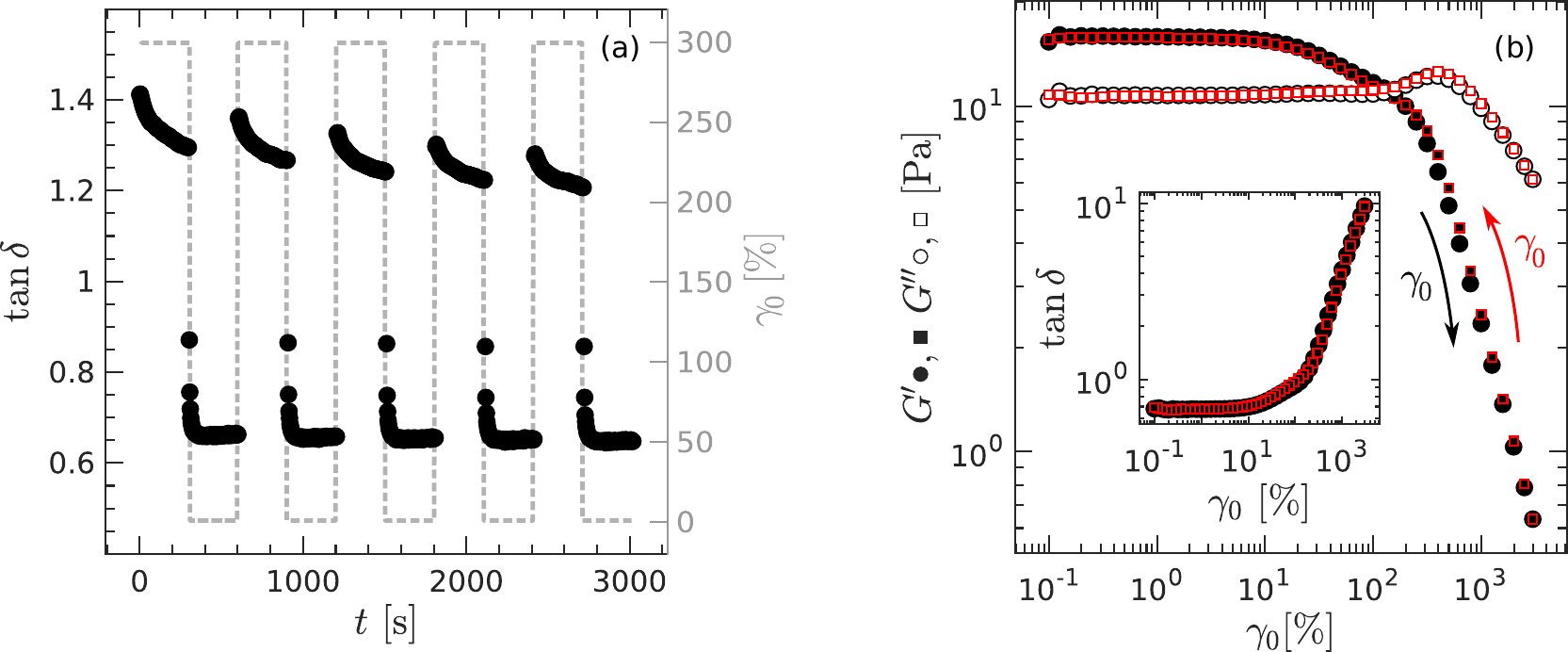}
\caption{(a) Temporal evolution of the loss factor $\tan \delta$ of a CMC gel under oscillatory shear at $f=1$~Hz during 5 steps of 300~s alternating large strain amplitude of 300\% and small strain amplitude of 1\%. 
The loss factor is represented in black (left axis) while the strain amplitude $\gamma_0$ is represented in gray (right axis). (b) Elastic and viscous moduli measured at $f=1$~Hz upon increasing the strain amplitude $\gamma_0$ (black circles) and subsequently decreasing $\gamma_0$ (red squares). The strain amplitude is varied logarithmically between 0.01\% and 3000\% with 10 steps per decade and each strain amplitude is applied for 3 cycles. Experiments conducted on a CMC gel prepared with $c_{\rm CMC}=3\%$ and $\mathrm{pH}=2.2$.}\label{fig:reversibility}
\end{figure*}

In this section, we examine the non-linear rheological response of acid-induced CMC gels. We first briefly test the reversibility of the shear-induced solid-to-liquid transition of a 3\% gel prepared at $\mathrm{pH}=2.2$, by imposing a series of consecutive large amplitude oscillatory shear (LAOS)  of amplitude $\gamma_0=300$\% for 300~s followed by a period of rest at small amplitude oscillatory shear (SAOS) $\gamma_0=1$\% for 300~s. Both steps are conducted at frequency $f=1$~Hz. The results are illustrated in Fig.~\ref{fig:reversibility}(a) where we report the loss factor $\tan \delta=G''/G'$ as a function of time over 5 successive LAOS and SAOS cycles. 

Under LAOS, the CMC gel flows as a viscous liquid ($\tan \delta >1$) and exhibits a time-dependent response [Fig.~\ref{fig:reversibility}(a)]. The loss factor decreases during the $300~\rm s$ of the test, without reaching a steady state. In practice, examining the individual evolution of $G'$ and $G''$ shows that $G''$ is roughly constant, whereas $G'$ increases [see Fig.~S2(a) in the Supplementary Information]. These observations show that acid-induced CMC gels flow like liquids under large amplitude oscillations, while their microstructure keeps rearranging under shear over a timescale larger than $300~\rm s$, most likely due to the association/dissociation dynamics of hydrophobic patches.

Upon flow cessation and switching to SAOS, the loss factor displays a rapid drop with $\tan \delta <1$ associated with the sol-gel transition before reaching a quasi steady-state in about $20~\rm s$ [Fig.~\ref{fig:reversibility}(a)]. The crossover of $G'$ and $G''$ occurs within seconds, while $G'(t)$ subsequently shows a mild overshoot with a maximum at about $50~\rm s$ [see Fig.~S2(b)]. These observations suggest that the percolated network responsible for the gel elasticity forms rapidly and is followed by \textit{local} and/or cooperative microstructural rearrangements leading to the slow relaxation of $G'$. Repeating these cycles of LAOS and SAOS reveals that (i)~acid-induced CMC gels reversibly yield under external shear and reform upon flow cessation, and (ii)~that the state reached upon flow cessation poorly depends on the shear history experienced by the sample.

To further illustrate the absence of memory in the rheological response of acid-induced CMC gels, we perform a strain-sweep experiment on the same 3\% CMC gel first by increasing the strain amplitude $\gamma_0$ before immediately decreasing it. The result is shown in Fig.~\ref{fig:reversibility}(b). Increasing the strain amplitude at $f=1~\rm Hz$ by discrete steps spaced logarithmically between 0.1\% to 3000\% leads to a shear-induced yielding transition [Fig.~\ref{fig:reversibility}(b)]. In practice, the elastic modulus $G'$ is constant in the linear deformation regime, and decreases monotonically beyond a strain amplitude of about 10\%. Conversely, the viscous modulus $G''$ first remains constant and smaller than $G'$, then crosses $G'$ at a critical strain $\gamma_y$ of about 150\%, before showing a maximum at $\gamma \simeq 400\%$ followed by a sharp decrease. The overshoot in $G''$ is referred to as the ``Payne effect" in filled polymer composites \cite{Xu:2019,Fan:2019}, while the overall response is classified as Type III, and commonly observed in a broad range of colloidal particulate and polymer gels \cite{Hyun:2002}. Interestingly, performing a decreasing ramp of strain following an identical strain path as during the increasing ramp leads to the very same viscoelastic response of the gel, confirming the reversibility of the shear-induced solid-to-liquid transition, as already demonstrated by alternating periods of LAOS and SAOS. 

\subsection{Intra-cycle analysis of oscillatory shear experiments}

\begin{figure}[!p]
\centering
\includegraphics[width = 0.5\linewidth]{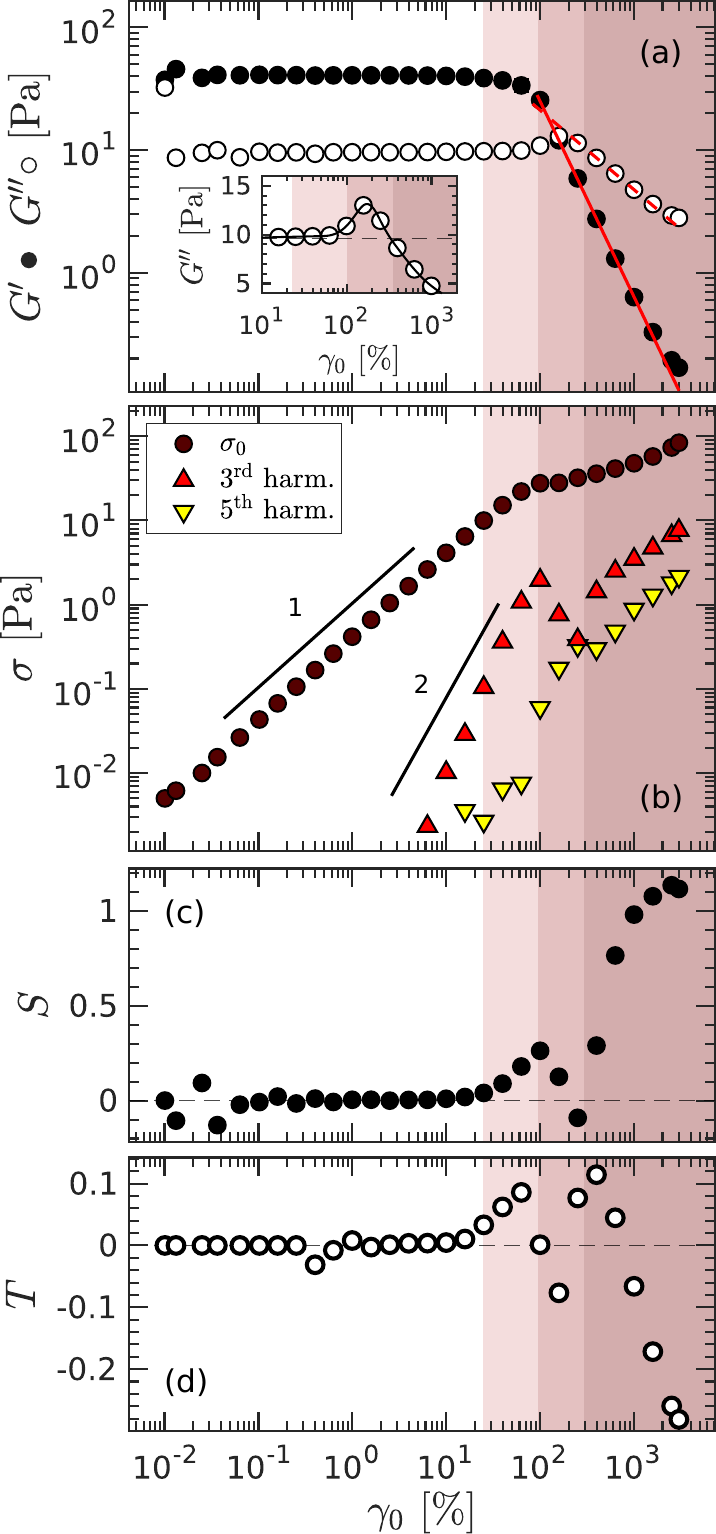}
\caption{Intra-cycle analysis of an oscillatory shear experiment performed on a $3\%$ CMC gel at $\mathrm{pH}=0.6$. (a) Viscoelastic moduli $G'$ ($\bullet$) and $G''$ $(\circ)$ vs.~strain amplitude $\gamma_0$. The red lines show the best power-law fit of the data beyond the crossover of $G'$ and $G''$, i.e., $G'\sim \gamma_0^{-\nu'}$ and $G''\sim \gamma_0^{-\nu''}$, with $\nu'=1.6$ and $\nu''=0.6$. Inset: zoom on the overshoot of $G''$ near the yield point defined by the crossover of $G'$ and $G''$ in the main figure. The black curve corresponds to the spline interpolation used to estimate the maximum of $G''$. (b) Stress amplitude $\sigma_0$ as well as third and fifth harmonics of the stress response vs.~strain amplitude $\gamma_0$. Black lines emphasize a linear and a quadratic increase. (c) Strain-stiffening ratio $S=(G'_L-G'_M)/G'_L$, with $G'_M=(d\sigma /d\gamma)\mid_{\gamma_0}$ the minimum-strain modulus, and $G'_L=(\sigma/\gamma)\mid_{\gamma=\pm \gamma_0}$ the large-strain modulus. 
(d) Shear-thickening ratio $T=(\eta'_L-\eta'_M)/\eta'_L$. 
In (c) and (d), the horizontal dashed line highlights $S=T=0$. In (a)--(d), the non-linear response is divided into three regions of strain highlighted by vertical stripes. The first region ($\gamma_0 \gtrsim 25\%$) marks the onset of non-linearity based on the intra-cycle analysis (defined by $S>0$ and $T>0$). The second region ($\gamma_0 \gtrsim 100\%$) marks the onset of non-linearity based on the strain-dependence of the stress amplitude $\sigma(\gamma_0)$. The third region ($\gamma_0 \gtrsim 300\%$) corresponds to the fully-developed non-linear response beyond the minimum of $S$ (and maximum of $T$).\label{fig:LAOStrain}}
\end{figure}

In this section, we further quantify the shear-induced yielding of acid-induced CMC gels upon increasing the strain amplitude. More precisely, a strain sweep is performed by ramping up the strain amplitude $\gamma_0$ from 0.01\% to 3000\% at fixed frequency $\omega = 2\pi ~\rm rad.s^{-1}$. The evolution of $G'$ and $G''$ vs.~the strain amplitude $\gamma_0$ is shown in Fig.~\ref{fig:LAOStrain}(a) together with the stress amplitude $\sigma(\gamma_0$) in Fig.~\ref{fig:LAOStrain}(b). The stress grows linearly up to the crossover of $G'$ and $G''$, which defines the yield strain $\gamma_y \simeq 200\%$ and the yield stress $\sigma_y \simeq 40~\rm Pa$. Additionally, we report the growth of harmonics in the stress signal: both the third and fifth harmonics grow quadratically with the strain amplitude up to the yield point. At the yield point, $G''$ shows a maximum [see inset in Fig.~\ref{fig:LAOStrain}(a)], while for $\gamma_0 >\gamma_y$ both $G'$ and $G''$ display a power-law decay highlighted with red lines in Fig.~\ref{fig:LAOStrain}(a). In practice, $G' \sim \gamma_0^{-\nu'}$ and $G'' \sim \gamma_0^{-\nu''}$, with $\nu'=1.6$ and $\nu''=0.6$, such that $\nu'/\nu''\simeq 2.7$. Remarkably, these values contrast with predictions from Mode Coupling Theory (MCT) following which, one should expect $\nu''\simeq 0.9$ and $\nu'=2\nu''$ \cite{Miyazaki:2006}. Such a discrepancy suggests that the shear-induced rearrangements of the polymer network, especially at the locus of hydrophobic patches might play a key role in the yielding process since MCT does not take into account local rearrangements of particles occurring via the crossing of activated barriers. Yet, acid-induced CMC gels are not an isolated case, for the prediction $\nu'=2\nu''$ has been reported to fail in several experimental studies on colloidal gels \cite{Sudreau:2022} and denser suspensions of soft repulsive particles, i.e., colloidal glasses \cite{Helgeson:2007,Carrier:2009,Koumakis:2012}.

\begin{figure*}[!h]
\centering
\includegraphics[width = 0.8\linewidth]{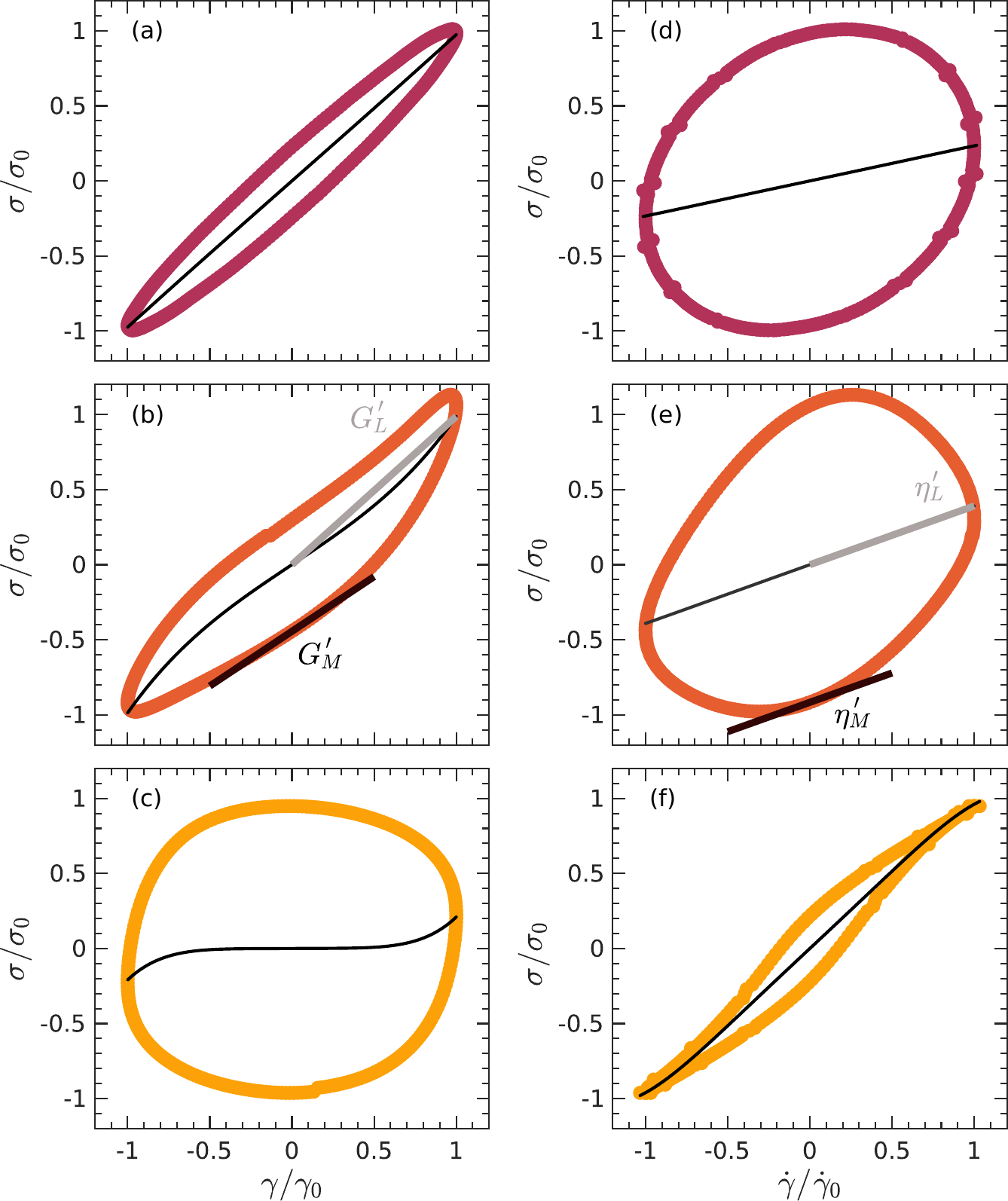}
\caption{Selected waveforms of the stress response to oscillatory strain deformations of increasing amplitude for a $3\%$ CMC gel at $\mathrm{pH}=0.6$ under various strain amplitudes: (a,d) $\gamma_0=10\%$, (b,e) 100\% , and (c,f) 1000\%. The normalized stress $\sigma/\sigma_0$ is reported as a function of either the normalized strain $\gamma/\gamma_0$ in (a), (b), and (c), or the normalized shear rate $\dot \gamma/\dot \gamma_0$, with $\dot \gamma_0=\omega \gamma_0$ in (d), (e), and (f). The black curves show the elastic (resp. viscous) part of the stress response in (a), (b), and (c) [resp. (d), (e), and (f)]. In (b), we define the elastic modulus at the minimum strain $G'_M=(d\sigma /d\gamma)\mid_{\gamma=0}$, and the elastic modulus at the largest strain $G'_L=(\sigma/\gamma)\mid_{\gamma= \gamma_0}$. In (e), we define the minimum rate dynamic viscosity $\eta'_M=(d\sigma/d\dot \gamma)\mid_{\dot \gamma=0}$ and the large rate dynamic viscosity $\eta'_L=(\sigma/\dot \gamma)\mid_{\dot \gamma=\dot \gamma_0}$. These moduli and viscosities are used to compute the parameters $S$ and $T$ shown in Fig.~\ref{fig:LAOStrain}(c) and \ref{fig:LAOStrain}(d).}\label{fig:lissajous}
\end{figure*}

To provide a more detailed picture of the yielding transition, we turn to the intra-cycle analysis of the above strain sweep experiments. Typical waveforms for the intra-cycle response of an acid-induced CMC hydrogel are reported in Fig.~\ref{fig:lissajous} as Lissajous-Bowditch (LB) plots that parametrically show the measured instantaneous stress against the imposed instantaneous strain $\gamma$ and strain rate $\dot \gamma$ within a period of oscillation \cite{Cho:2005}. We report LB plots for three different strain amplitudes along the strain sweep experiment, i.e., $\gamma_0=10\%$, 100\%, and 1000\%. For increasing strain amplitude, this representation illustrates very well the growth of harmonics in the stress signal and the transition from the linear regime, i.e., $\sigma (\gamma)$ close to a straight line and $\sigma (\dot \gamma)$ close to a circle, to the non-linear regime where $\sigma (\gamma)$ and $\sigma (\dot \gamma)$ display more complex shapes. In order to quantify the intra-cyle evolution of the sample response, we use two additional elastic moduli introduced in ref.~\cite{Ewoldt:2008}, which are the elastic modulus computed at the minimum strain within the cycle, i.e, $G'_M=(\partial \sigma /\partial \gamma)\mid_{\gamma=0}$, and the elastic modulus computed at the largest strain applied to the sample, i.e., $G'_L=(\sigma/\gamma)\mid_{\gamma= \gamma_0}$ [see Fig.~\ref{fig:lissajous}(b)]. The gel response within an oscillation is well captured by the strain-stiffening ratio $S$, a dimensionless quantity defined as $S=(G'_L-G'_M)/G'_L$, where $S>0$ indicates intra-cycle strain-stiffening, whereas $S<0$ corresponds to an intra-cycle strain-softening behavior. The same approach can be used on the waveform $\sigma(\dot \gamma)$ to define the minimum rate dynamic viscosity $\eta'_M=(d\sigma/d\dot \gamma)\mid_{\dot \gamma=0}$ and the large rate dynamic viscosity $\eta'_L=(\sigma/\dot \gamma)\mid_{\dot \gamma=\dot \gamma_0}$ with $\dot \gamma_0=\omega \gamma_0$ [see Fig.~\ref{fig:lissajous}(e)]. The corresponding dimensionless ratio, i.e., the shear-thickening ratio, is defined as $T=(\eta'_L-\eta'_M)/\eta'_L$, where $T>0$ implies intra-cycle shear-thickening, whereas $T<0$ corresponds to an intra-cycle shear-thinning behavior.

The parameters $S$ and $T$ are computed along the oscillatory strain ramp shown in Fig.~\ref{fig:LAOStrain}(a) and reported in Figs.~\ref{fig:LAOStrain}(c) and \ref{fig:LAOStrain}(d), respectively. At low strain amplitude, $S=T=0$; this linear regime extends up to $\gamma_0 \simeq 25\%$, which defines the onset of the non-linear response beyond which $S>0$ and $T>0$ (see light pink vertical stripe in Fig.~\ref{fig:LAOStrain}). Such intra-cycle strain-hardening and intra-cycle shear-thickening responses suggest that the polymer network responds non-linearly to the increasing deformation, although the increase in stress amplitude remains linear [Fig.~\ref{fig:LAOStrain}(b)]. Both parameters $S$ and $T$ increase up to a maximum reached at $\gamma_0=100\%$, beyond which $\sigma(\gamma_0)$ shows a sublinear increase. Then, the parameters $S$ and $T$ drop abruptly to become negative at the yield point $\gamma=\gamma_y\simeq 200\%$ defined by the crossover of $G'$ and $G''$ (see second vertical stripe in Fig.~\ref{fig:LAOStrain}). The gel response is both intra-cycle strain-softening and shear-thinning, which most likely reflects the loss of integrity of the gel and its failure due to the unbinding of a sufficiently large number of hydrophobic patches (``pull-out events"). Finally, for  $\gamma_0 \gtrsim 300\%$, the looser gel concomitantly displays intra-cycle shear-thinning ($T<0$) and strain-hardening ($S>0$) responses, being characteristic of the CMC strands alignment in the flow direction (see dark pink vertical stripe in Fig.~\ref{fig:LAOStrain}). While the former observation originates from the loss of topological interactions between the strands due to their preferential orientation, the latter is a direct consequence of their finite extensibility \cite{doi1979dynamics,nebouy2021flow}. 

\subsection{Impact of the pH on the non-linear features}

\begin{figure*}[!t]
\centering
\includegraphics[width = 0.9\linewidth]{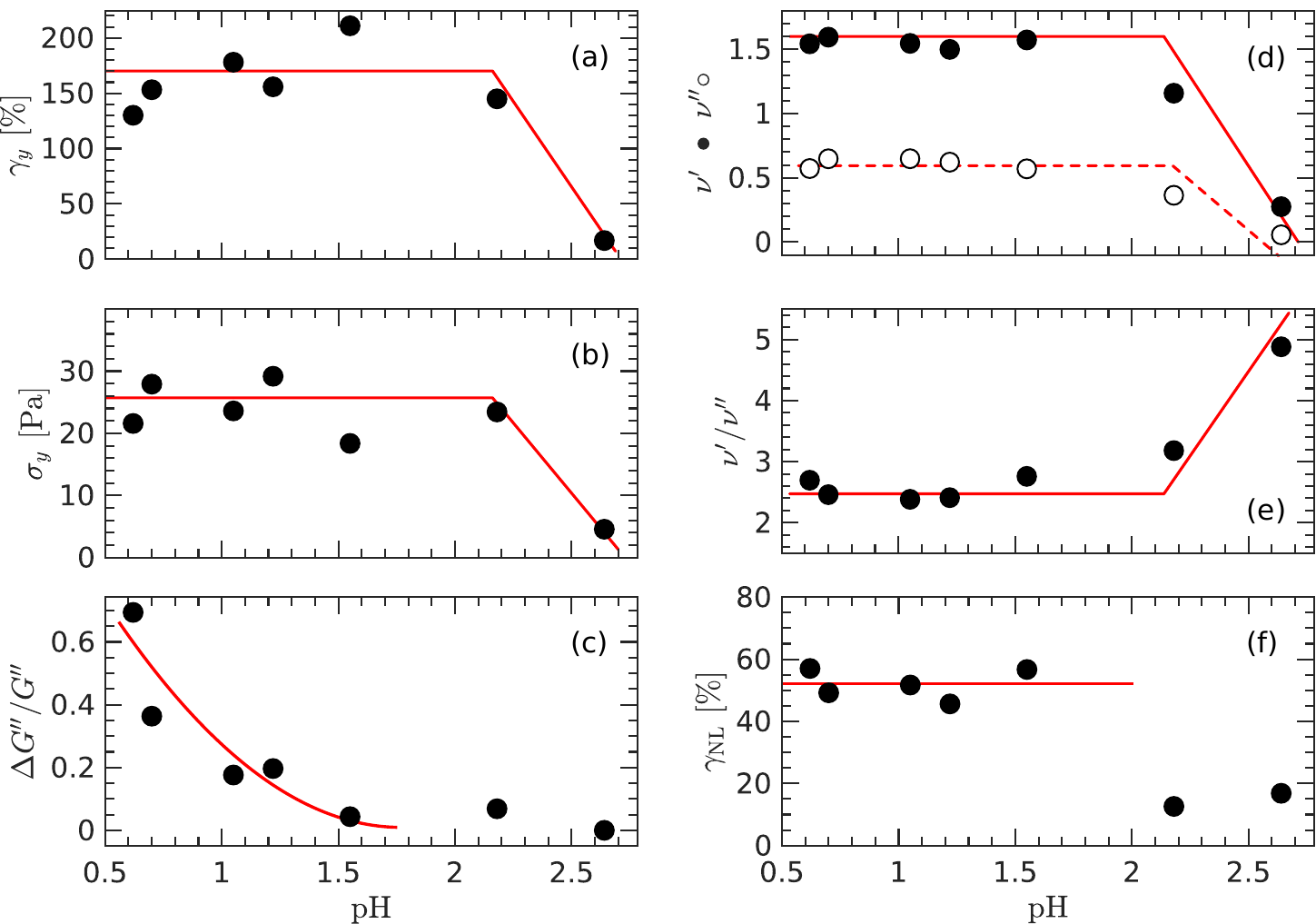}
\caption{Non-linear properties of a 3\% CMC gel reported as a function of the pH. (a) Yield strain $\gamma_y$ and (b) yield stress $\sigma_y$ defined by the crossover of $G'$ and $G''$ [see Fig.~\ref{fig:LAOStrain}(a)]. (c) Normalized amplitude $\Delta G''/G''$ of the maximum in $G''$, where $\Delta G''$ is the difference between the maximum of $G''$ and its linear value measured in the low strain limit. A spline interpolation of $G''(\gamma_0)$ is used to precisely determine the maximum of $G''$, as depicted in the inset in Fig.~\ref{fig:LAOStrain}(a). (d) Exponents $\nu'$ and $\nu''$ characterizing the power-law decay of $G'$ and $G''$ beyond the yield point; (e) ratio $\nu'/\nu''$. (f) Onset of non-linearity $\gamma_{\rm NL}$. Red curves are guides for the eye. }\label{fig:pHonrheology}
\end{figure*}

In this last section, we examine the influence of pH on the nonlinear mechanical response of acid-induced CMC gels. We focus on a 3\% CMC gel and report the key features of the non-linear experiments in Figure~\ref{fig:pHonrheology}, namely the onset of non-linearity $\gamma_{\rm NL}$, the yield strain $\gamma_y$ and the yield stress $\sigma_y$, the normalized amplitude $\Delta G''/G''$ of the maximum in $G''$ at the yield point, and the exponents characterizing the power-law decay of $G'$ and $G''$ beyond the yield point. For $c_\text{CMC}=3\%$, the sol-gel transition occurs at $\mathrm{pH_c}\simeq 3$ (see Fig.~\ref{fig:PD}). Below $\rm pH_c$, the observables $\gamma_y$, $\sigma_y$, $\nu'$, and $\nu''$ increase for decreasing pH over one pH unit. Below $\mathrm{pH} \simeq 2.2$, all properties but the $G''$ overshoot remain constant: $\gamma_{NL} \simeq 55\%$, $\gamma_y \simeq 150\%$, and $\sigma_y \simeq 25~\rm Pa$, while $\nu'\simeq 1.6$, $\nu'' \simeq 0.6$, and $\nu'/\nu'' \simeq 2.7$. This observation suggests that most of the hydrophobic patches are bonded beyond $\mathrm{pH} \simeq 2.2$,
in agreement with results reported by neutron scattering and rheometry in ref.~\cite{Legrand:2024}.

However, the normalized amplitude of the $G''$ overshoot at the yield point keeps growing for decreasing pH. This indicates that the viscous dissipation becomes more and more pronounced as the pH of the CMC gel decreases. From a microscopic perspective, the overshoot in $G''$ is linked to plastic flow and originates in the \textit{continuous} transition from recoverable to unrecoverable strain accumulation \cite{Pham:2006,Donley:2020}. In that framework, the progressive increase of the amplitude in the $G''$ overshoot for decreasing pH can be interpreted as the relative decrease of the viscoelastic solid dissipation compared to the unrecoverable dissipation due to plastic flow \cite{Donley:2020}. This result strongly suggests that the yielding transition of acid-induced CMC gels becomes spatially more heterogeneous for decreasing pH, which remains to be tested by performing temporally and spatially-resolved measurements across the yielding transition \cite{Manneville:2008,SaintMichel2016}. 

\section{Conclusion}\label{sec:Conclusion}

We have shown that acidified aqueous suspensions of carboxymethylcellulose obey a time-temperature superposition principle across the sol-gel transition. In the sol phase, the viscoelastic spectrum displays a power-law frequency dependence that is well-captured by a fractional Maxwell model, while the viscoelastic spectrum is captured by a fractional Kelvin-Voigt model. In both cases, the activation energies extracted from the temperature dependence of the rescaling factors are compatible with the presence of hydrophobic patches along the polymer chain, driving the association of the CMC. While these two fractional models show two power-law exponents of similar values (i.e., $\kappa$ and $\alpha$), it remains unclear whether these two descriptions can be unified into a single model capturing the viscoelastic spectrum across the sol-gel transition, in the spirit of recent work on the sol-gel transition in aqueous dispersions of cellulose nanocrystals \cite{MorletDecarnin:2023}. 

Acid-induced CMC gels follow the rubber elasticity model and display a reversible shear-induced yielding transition, which follows a Type III response under strain sweep experiments. Beyond the yield point, the viscoelastic moduli display a decreasing power-law response for increasing strain amplitude. The exponents are not captured by MCT, which prompts one to obtain predictions from other classical theoretical approaches capturing the shear-induced solid-to-liquid transition. Promising options lie in recent continuum modeling such as the spatially-resolved fluidity model that encompasses non-local effects and was shown to quantitatively predict shear-induced fluidization scenarios \cite{Benzi:2023}, as well as the recent elasto-visco-plastic model proposed in refs.~\cite{Kamani:2021,Kamani:2024} that captures the salient features of LAOS.

To conclude, we believe the present work paves the way for a more systematic investigation of yielding in polymer gels formed by hydrophobic interactions. Future work should focus on identifying the features of the yielding process that are specific and/or affected by the very nature of the hydrophobic interactions.

\backmatter





\bmhead{Acknowledgments}

The authors thank M.~Peyla for preliminary experiments, A.~Crepet for the size exclusion chromatography test, as well as J.~Bauland, J.~Blin, and G.~Petekidis for fruitful discussions. This work was supported by the LABEX iMUST of the University of Lyon (ANR-10-LABX-0064), created within the ``Plan France 2030" set up by the French government and managed by the French National Research Agency (ANR).  


\section*{Declarations}

\textbf{Competing Interests} The authors declare that they have no conflicts of interest.
\\
\\

\noindent

\clearpage
\newpage
\setcounter{page}{1}
\setcounter{equation}{0}
\setcounter{figure}{0}
\global\def\thefigure{S\arabic{figure}}
\setcounter{table}{0}
\global\def\thetable{S\arabic{table}}

\begin{center}
    {\large\bf {\sc Supplementary information}}
\end{center}
\begin{center}
    {\large\bf Acid-induced carboxymethylcellulose hydrogels}
\end{center}

\subsection*{Aging of a 3\% CMC gel monitored by Time-resolved Mechanical Spectroscopy}

\begin{figure*}[!h]
\centering
\includegraphics[width = 0.9\linewidth]{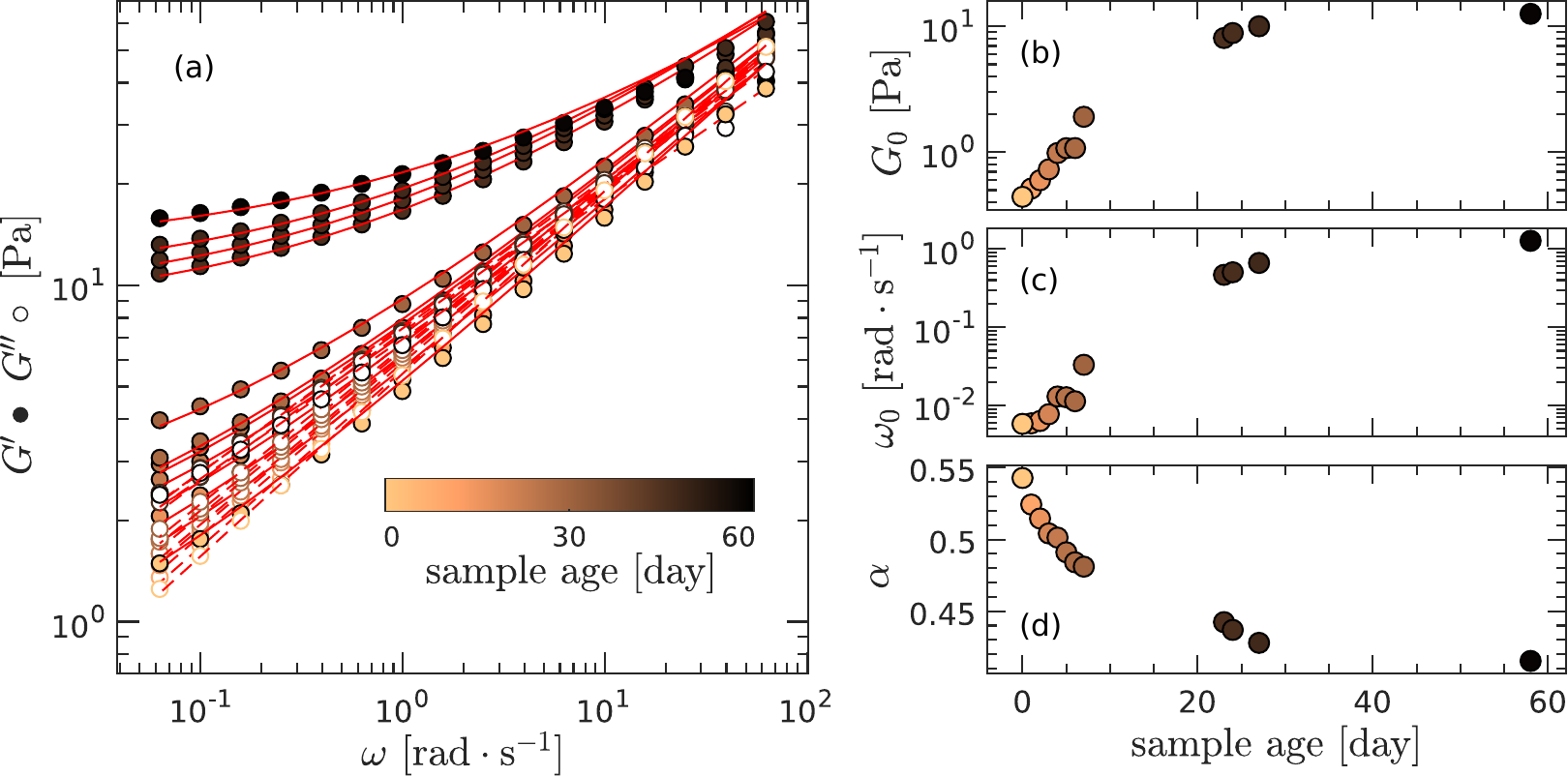}
\caption{Long-time evolution of the viscoelastic spectrum in the gel phase. (a) Viscoelastic moduli $G'$ ($\bullet$) and $G''$ ($\circ$) as a function of frequency $\omega$. The red curves are the best FKV fits [see Eq.~\eqref{eq:FKV} in the main text], whose parameters are plotted in (b-d). Colors code for the sample age as shown in the color bar. (b) Modulus scale $G_0$, (c) frequency scale $\omega_0$, and (d) exponent $\alpha$ as a function of the sample age, defined as the time elapsed since the preparation date. The samples are CMC gels at fixed composition from the same batch ($c_{\rm CMC}=3\%$ and $\mathrm{pH}=1.6$). Experiments performed at a strain amplitude $\gamma_0=1~\rm \%$ and fixed temperature $T=22^\circ \rm C$.}\label{fig:sample_age_gel}
\end{figure*}

In order to test the stability of acid-induced CMC gels, we have prepared a batch of acid-induced CMC gels with $c_\text{CMC}=3\%$, and $\mathrm{pH}=1.6$ that was left to age at room temperature. Samples were regularly extracted, and their linear viscoelastic spectrum was measured and fitted by a fractional Kelvin-Voigt model. The results are reported in Fig.~\ref{fig:sample_age_gel}. Over the course of two months, the gel viscoelastic properties are reinforced, as evidenced by the increase in $G_0$, while $\alpha$ decreases from $0.55$ down to $0.41$. We attribute these aging dynamics to the slow reorganization of the gel microstructure associated with the formation of new and/or more stable bonds between hydrophobic patches along the polymer chain.  

\clearpage

\subsection*{Viscoelastic moduli under LAOS and SAOS}

\begin{figure*}[!h]
\centering
\includegraphics[width = 0.9\linewidth]{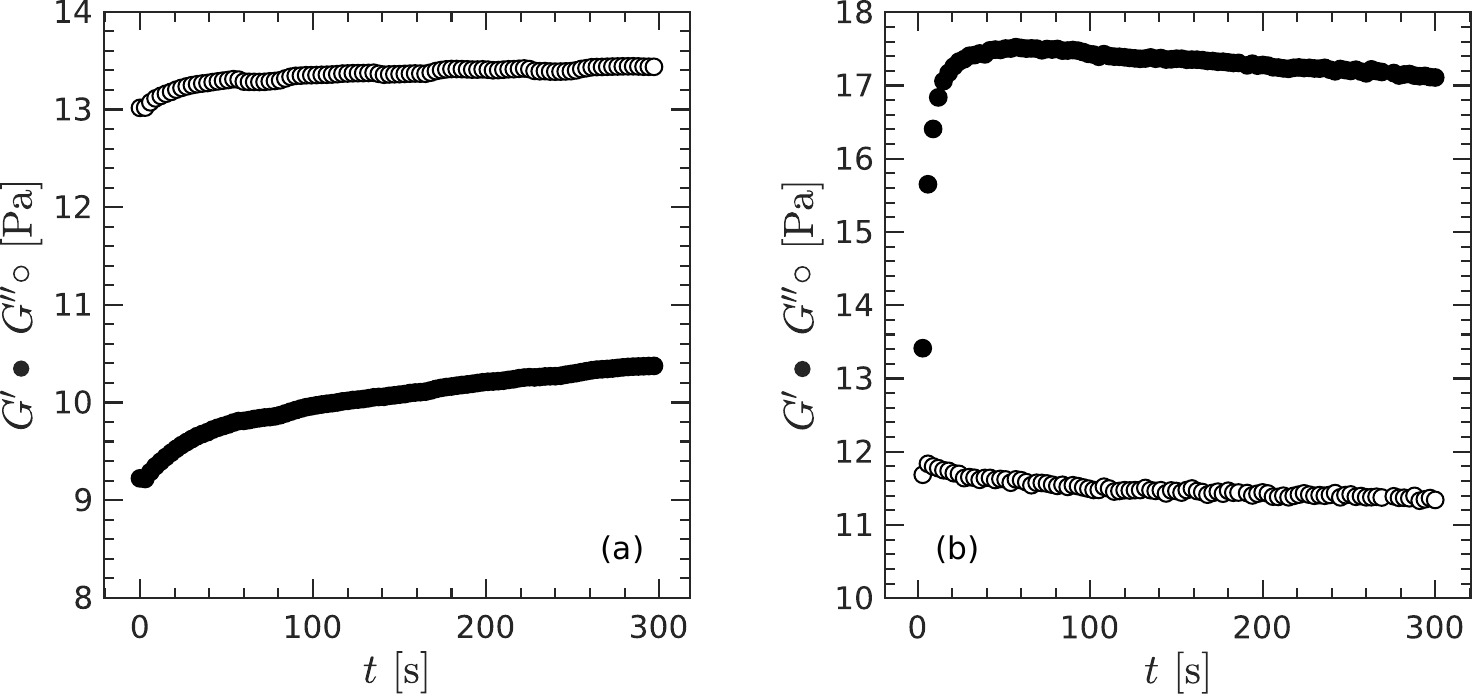}
\caption{Temporal evolution of the viscoelastic modulus $G'$ and $G''$ measured at $f=1~\rm Hz$ during (a) LAOS and (b) SAOS. 
These data correspond to steps one and two in Fig.~6 in the main text. Experiments were performed on a 3\% CMC gel prepared at $\mathrm{pH}=2.2$.  }\label{fig:reversibility_dynamic}
\end{figure*}

Here, we highlight the temporal evolution of the viscoelastic moduli during the experiment reported in Fig.~6 in the main text, namely large amplitude oscillations (LAOS) in Fig.~\ref{fig:reversibility_dynamic}(a) followed by small amplitude oscillations (SAOS) in Fig.~\ref{fig:reversibility_dynamic}(b). Under LAOS, the CMC gel displays a liquid-like response ($G'<G''$), with a constant viscous dissipation, whereas $G'$ slowly increases over 300~s. Upon flow cessation, the gel rapidly regains its elastic behavior with $G'>G''$. The elastic modulus $G'$ shows an overshoot, which suggests that the microstructure that forms upon the rapid gelation of the sample slowly relaxes due to thermal agitation, and/or some residual stresses trapped upon flow cessation.

\begin{thebibliography}{67}
\ifx \bisbn   \undefined \def \bisbn  #1{ISBN #1}\fi
\ifx \binits  \undefined \def \binits#1{#1}\fi
\ifx \bauthor  \undefined \def \bauthor#1{#1}\fi
\ifx \batitle  \undefined \def \batitle#1{#1}\fi
\ifx \bjtitle  \undefined \def \bjtitle#1{#1}\fi
\ifx \bvolume  \undefined \def \bvolume#1{\textbf{#1}}\fi
\ifx \byear  \undefined \def \byear#1{#1}\fi
\ifx \bissue  \undefined \def \bissue#1{#1}\fi
\ifx \bfpage  \undefined \def \bfpage#1{#1}\fi
\ifx \blpage  \undefined \def \blpage #1{#1}\fi
\ifx \burl  \undefined \def \burl#1{\textsf{#1}}\fi
\ifx \doiurl  \undefined \def \doiurl#1{\url{https://doi.org/#1}}\fi
\ifx \betal  \undefined \def \betal{\textit{et al.}}\fi
\ifx \binstitute  \undefined \def \binstitute#1{#1}\fi
\ifx \binstitutionaled  \undefined \def \binstitutionaled#1{#1}\fi
\ifx \bctitle  \undefined \def \bctitle#1{#1}\fi
\ifx \beditor  \undefined \def \beditor#1{#1}\fi
\ifx \bpublisher  \undefined \def \bpublisher#1{#1}\fi
\ifx \bbtitle  \undefined \def \bbtitle#1{#1}\fi
\ifx \bedition  \undefined \def \bedition#1{#1}\fi
\ifx \bseriesno  \undefined \def \bseriesno#1{#1}\fi
\ifx \blocation  \undefined \def \blocation#1{#1}\fi
\ifx \bsertitle  \undefined \def \bsertitle#1{#1}\fi
\ifx \bsnm \undefined \def \bsnm#1{#1}\fi
\ifx \bsuffix \undefined \def \bsuffix#1{#1}\fi
\ifx \bparticle \undefined \def \bparticle#1{#1}\fi
\ifx \barticle \undefined \def \barticle#1{#1}\fi
\bibcommenthead
\ifx \bconfdate \undefined \def \bconfdate #1{#1}\fi
\ifx \botherref \undefined \def \botherref #1{#1}\fi
\ifx \url \undefined \def \url#1{\textsf{#1}}\fi
\ifx \bchapter \undefined \def \bchapter#1{#1}\fi
\ifx \bbook \undefined \def \bbook#1{#1}\fi
\ifx \bcomment \undefined \def \bcomment#1{#1}\fi
\ifx \oauthor \undefined \def \oauthor#1{#1}\fi
\ifx \citeauthoryear \undefined \def \citeauthoryear#1{#1}\fi
\ifx \endbibitem  \undefined \def \endbibitem {}\fi
\ifx \bconflocation  \undefined \def \bconflocation#1{#1}\fi
\ifx \arxivurl  \undefined \def \arxivurl#1{\textsf{#1}}\fi
\csname PreBibitemsHook\endcsname

\bibitem{Zhang:2017}
\begin{barticle}
\bauthor{\bsnm{Zhang}, \binits{Y.S.}},
\bauthor{\bsnm{Khademhosseini}, \binits{A.}}:
\batitle{Advances in engineering hydrogels}.
\bjtitle{Science}
\bvolume{356}(\bissue{6337}),
\bfpage{3627}
(\byear{2017})
\end{barticle}
\endbibitem

\bibitem{Cao:2020}
\begin{barticle}
\bauthor{\bsnm{Cao}, \binits{Y.}},
\bauthor{\bsnm{Mezzenga}, \binits{R.}}:
\batitle{Design principles of food gels}.
\bjtitle{Nature Food}
\bvolume{1}(\bissue{2}),
\bfpage{106}--\blpage{118}
(\byear{2020})
\end{barticle}
\endbibitem

\bibitem{Rossow:2015}
\begin{botherref}
\oauthor{\bsnm{Rossow}, \binits{T.}},
\oauthor{\bsnm{Seiffert}, \binits{S.}}:
Supramolecular polymer networks: Preparation, properties, and potential.
Supramolecular Polymer Networks and Gels,
1--46
(2015)
\end{botherref}
\endbibitem

\bibitem{Khare:2021}
\begin{barticle}
\bauthor{\bsnm{Khare}, \binits{E.}},
\bauthor{\bsnm{Holten-Andersen}, \binits{N.}},
\bauthor{\bsnm{Buehler}, \binits{M.J.}}:
\batitle{Transition-metal coordinate bonds for bioinspired macromolecules with
  tunable mechanical properties}.
\bjtitle{Nat. Rev. Mater.}
\bvolume{6}(\bissue{5}),
\bfpage{421}--\blpage{436}
(\byear{2021})
\end{barticle}
\endbibitem

\bibitem{jiang2022magneto}
\begin{barticle}
\bauthor{\bsnm{Jiang}, \binits{L.}},
\bauthor{\bsnm{Griffiths}, \binits{P.}},
\bauthor{\bsnm{Balouet}, \binits{J.}},
\bauthor{\bsnm{Faure}, \binits{T.}},
\bauthor{\bsnm{Lyons}, \binits{R.}},
\bauthor{\bsnm{Fustin}, \binits{C.-A.}},
\bauthor{\bsnm{Baeza}, \binits{G.P.}}:
\batitle{Magneto-responsive nanocomposites with a metal--ligand supramolecular
  matrix}.
\bjtitle{Macromolecules}
\bvolume{55}(\bissue{10}),
\bfpage{3936}--\blpage{3947}
(\byear{2022})
\end{barticle}
\endbibitem

\bibitem{zhuge2017decoding}
\begin{barticle}
\bauthor{\bsnm{Zhuge}, \binits{F.}},
\bauthor{\bsnm{Hawke}, \binits{L.G.}},
\bauthor{\bsnm{Fustin}, \binits{C.-A.}},
\bauthor{\bsnm{Gohy}, \binits{J.-F.}},
\bauthor{\bsnm{Van~Ruymbeke}, \binits{E.}}:
\batitle{Decoding the linear viscoelastic properties of model telechelic
  metallo-supramolecular polymers}.
\bjtitle{J. Rheol.}
\bvolume{61}(\bissue{6}),
\bfpage{1245}--\blpage{1262}
(\byear{2017})
\end{barticle}
\endbibitem

\bibitem{cui2018linear}
\begin{barticle}
\bauthor{\bsnm{Cui}, \binits{G.}},
\bauthor{\bsnm{Boudara}, \binits{V.A.}},
\bauthor{\bsnm{Huang}, \binits{Q.}},
\bauthor{\bsnm{Baeza}, \binits{G.P.}},
\bauthor{\bsnm{Wilson}, \binits{A.J.}},
\bauthor{\bsnm{Hassager}, \binits{O.}},
\bauthor{\bsnm{Read}, \binits{D.J.}},
\bauthor{\bsnm{Mattsson}, \binits{J.}}:
\batitle{Linear shear and nonlinear extensional rheology of unentangled
  supramolecular side-chain polymers}.
\bjtitle{J. Rheol.}
\bvolume{62}(\bissue{5}),
\bfpage{1155}--\blpage{1174}
(\byear{2018})
\end{barticle}
\endbibitem

\bibitem{louhichi2017humidity}
\begin{barticle}
\bauthor{\bsnm{Louhichi}, \binits{A.}},
\bauthor{\bsnm{Jacob}, \binits{A.}},
\bauthor{\bsnm{Bouteiller}, \binits{L.}},
\bauthor{\bsnm{Vlassopoulos}, \binits{D.}}:
\batitle{Humidity affects the viscoelastic properties of supramolecular living
  polymers}.
\bjtitle{J. Rheol.}
\bvolume{61}(\bissue{6}),
\bfpage{1173}--\blpage{1182}
(\byear{2017})
\end{barticle}
\endbibitem

\bibitem{baeza2016network}
\begin{barticle}
\bauthor{\bsnm{Baeza}, \binits{G.P.}},
\bauthor{\bsnm{Dessi}, \binits{C.}},
\bauthor{\bsnm{Costanzo}, \binits{S.}},
\bauthor{\bsnm{Zhao}, \binits{D.}},
\bauthor{\bsnm{Gong}, \binits{S.}},
\bauthor{\bsnm{Alegria}, \binits{A.}},
\bauthor{\bsnm{Colby}, \binits{R.H.}},
\bauthor{\bsnm{Rubinstein}, \binits{M.}},
\bauthor{\bsnm{Vlassopoulos}, \binits{D.}},
\bauthor{\bsnm{Kumar}, \binits{S.K.}}:
\batitle{Network dynamics in nanofilled polymers}.
\bjtitle{Nat. Commun.}
\bvolume{7}(\bissue{1}),
\bfpage{11368}
(\byear{2016})
\end{barticle}
\endbibitem

\bibitem{bouteiller2007assembly}
\begin{bbook}
\bauthor{\bsnm{Bouteiller}, \binits{L.}}:
In: \beditor{\bsnm{Binder}, \binits{W.}} (ed.)
\bbtitle{Assembly via Hydrogen Bonds of Low Molar Mass Compounds into
  Supramolecular Polymers},
pp. \bfpage{79}--\blpage{112}.
\bpublisher{Springer},
\blocation{Berlin, Heidelberg}
(\byear{2007})
\end{bbook}
\endbibitem

\bibitem{shabbir2017nonlinear}
\begin{barticle}
\bauthor{\bsnm{Shabbir}, \binits{A.}},
\bauthor{\bsnm{Huang}, \binits{Q.}},
\bauthor{\bsnm{Baeza}, \binits{G.P.}},
\bauthor{\bsnm{Vlassopoulos}, \binits{D.}},
\bauthor{\bsnm{Chen}, \binits{Q.}},
\bauthor{\bsnm{Colby}, \binits{R.H.}},
\bauthor{\bsnm{Alvarez}, \binits{N.J.}},
\bauthor{\bsnm{Hassager}, \binits{O.}}:
\batitle{Nonlinear shear and uniaxial extensional rheology of
  polyether-ester-sulfonate copolymer ionomer melts}.
\bjtitle{J. Rheol.}
\bvolume{61}(\bissue{6}),
\bfpage{1279}--\blpage{1289}
(\byear{2017})
\end{barticle}
\endbibitem

\bibitem{chen2013ionomer}
\begin{barticle}
\bauthor{\bsnm{Chen}, \binits{Q.}},
\bauthor{\bsnm{Tudryn}, \binits{G.J.}},
\bauthor{\bsnm{Colby}, \binits{R.H.}}:
\batitle{Ionomer dynamics and the sticky rouse model}.
\bjtitle{J. Rheol.}
\bvolume{57}(\bissue{5}),
\bfpage{1441}--\blpage{1462}
(\byear{2013})
\end{barticle}
\endbibitem

\bibitem{mei2022anion}
\begin{barticle}
\bauthor{\bsnm{Mei}, \binits{W.}},
\bauthor{\bsnm{Yu}, \binits{D.}},
\bauthor{\bsnm{George}, \binits{C.}},
\bauthor{\bsnm{Madsen}, \binits{L.A.}},
\bauthor{\bsnm{Hickey}, \binits{R.J.}},
\bauthor{\bsnm{Colby}, \binits{R.H.}}:
\batitle{Anion chemical composition of poly (ethylene oxide)-based
  sulfonylimide and sulfonate lithium ionomers controls ion aggregation and
  conduction}.
\bjtitle{J. Mater. Chem. C}
\bvolume{10}(\bissue{39}),
\bfpage{14569}--\blpage{14579}
(\byear{2022})
\end{barticle}
\endbibitem

\bibitem{Tuncaboylu:2012}
\begin{barticle}
\bauthor{\bsnm{Tuncaboylu}, \binits{D.C.}},
\bauthor{\bsnm{Argun}, \binits{A.}},
\bauthor{\bsnm{Sahin}, \binits{M.}},
\bauthor{\bsnm{Sari}, \binits{M.}},
\bauthor{\bsnm{Okay}, \binits{O.}}:
\batitle{Structure optimization of self-healing hydrogels formed via
  hydrophobic interactions}.
\bjtitle{Polymer}
\bvolume{53}(\bissue{24}),
\bfpage{5513}--\blpage{5522}
(\byear{2012})
\end{barticle}
\endbibitem

\bibitem{Cerny:2007}
\begin{barticle}
\bauthor{\bsnm{{\v{C}}ern{\`y}}, \binits{J.}},
\bauthor{\bsnm{Hobza}, \binits{P.}}:
\batitle{Non-covalent interactions in biomacromolecules}.
\bjtitle{Phys. Chem. Chem. Phys.}
\bvolume{9}(\bissue{39}),
\bfpage{5291}--\blpage{5303}
(\byear{2007})
\end{barticle}
\endbibitem

\bibitem{Wang:2017}
\begin{barticle}
\bauthor{\bsnm{Wang}, \binits{W.}},
\bauthor{\bsnm{Zhang}, \binits{Y.}},
\bauthor{\bsnm{Liu}, \binits{W.}}:
\batitle{Bioinspired fabrication of high strength hydrogels from non-covalent
  interactions}.
\bjtitle{Prog. Polym. Sci.}
\bvolume{71},
\bfpage{1}--\blpage{25}
(\byear{2017})
\end{barticle}
\endbibitem

\bibitem{Heinze:1999}
\begin{barticle}
\bauthor{\bsnm{Heinze}, \binits{T.}},
\bauthor{\bsnm{Pfeiffer}, \binits{K.}}:
\batitle{Studies on the synthesis and characterization of
  carboxymethylcellulose}.
\bjtitle{Angew. Makromolek. Chem.}
\bvolume{266}(\bissue{1}),
\bfpage{37}--\blpage{45}
(\byear{1999})
\end{barticle}
\endbibitem

\bibitem{Debutts:1957}
\begin{barticle}
\bauthor{\bsnm{DeButts}, \binits{E.}},
\bauthor{\bsnm{Hudy}, \binits{J.}},
\bauthor{\bsnm{Elliott}, \binits{J.}}:
\batitle{Rheology of sodium carboxymethylcellulose solutions}.
\bjtitle{Ind. Eng. Chem.}
\bvolume{49}(\bissue{1}),
\bfpage{94}--\blpage{98}
(\byear{1957})
\end{barticle}
\endbibitem

\bibitem{Lopez:2018}
\begin{barticle}
\bauthor{\bsnm{Lopez}, \binits{C.G.}},
\bauthor{\bsnm{Colby}, \binits{R.H.}},
\bauthor{\bsnm{Cabral}, \binits{J.T.}}:
\batitle{Electrostatic and hydrophobic interactions in nacmc aqueous solutions:
  Effect of degree of substitution}.
\bjtitle{Macromolecules}
\bvolume{51}(\bissue{8}),
\bfpage{3165}--\blpage{3175}
(\byear{2018})
\end{barticle}
\endbibitem

\bibitem{Elliot:1974}
\begin{barticle}
\bauthor{\bsnm{Elliot}, \binits{J.H.}},
\bauthor{\bsnm{Ganz}, \binits{A.}}:
\batitle{Some rheological properties of sodium carboxymethylcellulose solutions
  and gels}.
\bjtitle{Rheol. Acta}
\bvolume{13}(\bissue{4}),
\bfpage{670}--\blpage{674}
(\byear{1974})
\end{barticle}
\endbibitem

\bibitem{Barba:2002}
\begin{barticle}
\bauthor{\bsnm{Barba}, \binits{C.}},
\bauthor{\bsnm{Montan{\'e}}, \binits{D.}},
\bauthor{\bsnm{Farriol}, \binits{X.}},
\bauthor{\bsnm{Desbri{\`e}res}, \binits{J.}},
\bauthor{\bsnm{Rinaudo}, \binits{M.}}:
\batitle{Synthesis and characterization of carboxymethylcelluloses from
  non-wood pulps ii. rheological behavior of {CMC} in aqueous solution}.
\bjtitle{Cellulose}
\bvolume{9}(\bissue{3}),
\bfpage{327}--\blpage{335}
(\byear{2002})
\end{barticle}
\endbibitem

\bibitem{Lopez:2021}
\begin{barticle}
\bauthor{\bsnm{Lopez}, \binits{C.G.}},
\bauthor{\bsnm{Richtering}, \binits{W.}}:
\batitle{Oscillatory rheology of carboxymethyl cellulose gels: Influence of
  concentration and {pH}}.
\bjtitle{Carbohydr. Polym.}
\bvolume{267},
\bfpage{118117}
(\byear{2021})
\end{barticle}
\endbibitem

\bibitem{Dogsa:2014}
\begin{barticle}
\bauthor{\bsnm{Dogsa}, \binits{I.}},
\bauthor{\bsnm{Tom{\v{s}}i{\v{c}}}, \binits{M.}},
\bauthor{\bsnm{Orehek}, \binits{J.}},
\bauthor{\bsnm{Benigar}, \binits{E.}},
\bauthor{\bsnm{Jamnik}, \binits{A.}},
\bauthor{\bsnm{Stopar}, \binits{D.}}:
\batitle{Amorphous supramolecular structure of carboxymethyl cellulose in
  aqueous solution at different ph values as determined by rheology, small
  angle x-ray and light scattering}.
\bjtitle{Carbohydr. Polym.}
\bvolume{111},
\bfpage{492}--\blpage{504}
(\byear{2014})
\end{barticle}
\endbibitem

\bibitem{Legrand:2024}
\begin{barticle}
\bauthor{\bsnm{Legrand}, \binits{G.}},
\bauthor{\bsnm{Baeza}, \binits{G.P.}},
\bauthor{\bsnm{Peyla}, \binits{M.}},
\bauthor{\bsnm{Porcar}, \binits{L.}},
\bauthor{\bsnm{Fernandez-de-Alba}, \binits{C.}},
\bauthor{\bsnm{Manneville}, \binits{S.}},
\bauthor{\bsnm{Divoux}, \binits{T.}}:
\batitle{Acid-induced gelation of carboxymethylcellulose solutions}.
\bjtitle{ACS Macro Lett.}
\bvolume{13},
\bfpage{234}--\blpage{239}
(\byear{2024})
\end{barticle}
\endbibitem

\bibitem{Durig:1950}
\begin{barticle}
\bauthor{\bsnm{D{\"u}rig}, \binits{G.}},
\bauthor{\bsnm{Banderet}, \binits{A.}}:
\batitle{Sur la structure des solutions aqueuses de carboxymethylcellulose}.
\bjtitle{Helv. Chim. Acta}
\bvolume{33}(\bissue{4}),
\bfpage{1106}--\blpage{1118}
(\byear{1950})
\end{barticle}
\endbibitem

\bibitem{Hermans:1965}
\begin{barticle}
\bauthor{\bsnm{Hermans~Jr}, \binits{J.}}:
\batitle{Investigation of the elastic properties of the particle network in
  gelled solutions of hydrocolloids. i. carboxymethyl cellulose}.
\bjtitle{J. Polym. Sci., Part A: Gen. Pap.}
\bvolume{3}(\bissue{5}),
\bfpage{1859}--\blpage{1868}
(\byear{1965})
\end{barticle}
\endbibitem

\bibitem{Liebert:2001}
\begin{barticle}
\bauthor{\bsnm{Liebert}, \binits{T.F.}},
\bauthor{\bsnm{Heinze}, \binits{T.J.}}:
\batitle{Exploitation of reactivity and selectivity in cellulose
  functionalization using unconventional media for the design of products
  showing new superstructures}.
\bjtitle{Biomacromolecules}
\bvolume{2}(\bissue{4}),
\bfpage{1124}--\blpage{1132}
(\byear{2001})
\end{barticle}
\endbibitem

\bibitem{Liebert:2005}
\begin{bchapter}
\bauthor{\bsnm{Liebert}, \binits{T.}},
\bauthor{\bsnm{Hornig}, \binits{S.}},
\bauthor{\bsnm{Hesse}, \binits{S.}},
\bauthor{\bsnm{Heinze}, \binits{T.}}:
\bctitle{Microscopic visualization of nanostructures of cellulose derivatives}.
In: \bbtitle{Macromolecular Symposia},
vol. \bseriesno{223},
pp. \bfpage{253}--\blpage{266}
(\byear{2005}).
\bcomment{Wiley Online Library}
\end{bchapter}
\endbibitem

\bibitem{Bonn:2017}
\begin{barticle}
\bauthor{\bsnm{Bonn}, \binits{D.}},
\bauthor{\bsnm{Denn}, \binits{M.M.}},
\bauthor{\bsnm{Berthier}, \binits{L.}},
\bauthor{\bsnm{Divoux}, \binits{T.}},
\bauthor{\bsnm{Manneville}, \binits{S.}}:
\batitle{Yield stress materials in soft condensed matter}.
\bjtitle{Rev. Mod. Phys.}
\bvolume{89},
\bfpage{035005}
(\byear{2017})
\end{barticle}
\endbibitem

\bibitem{Viasnoff:2002}
\begin{barticle}
\bauthor{\bsnm{Viasnoff}, \binits{V.}},
\bauthor{\bsnm{Lequeux}, \binits{F.}}:
\batitle{Rejuvenation and overaging in a colloidal glass under shear}.
\bjtitle{Phys. Rev. Lett.}
\bvolume{89},
\bfpage{065701}
(\byear{2002})
\end{barticle}
\endbibitem

\bibitem{Joshi:2018}
\begin{barticle}
\bauthor{\bsnm{Joshi}, \binits{Y.M.}},
\bauthor{\bsnm{Petekidis}, \binits{G.}}:
\batitle{Yield stress fluids and ageing}.
\bjtitle{Rheol. Acta}
\bvolume{57},
\bfpage{521}--\blpage{549}
(\byear{2018})
\end{barticle}
\endbibitem

\bibitem{Winter:1997}
\begin{barticle}
\bauthor{\bsnm{Winter}, \binits{H.}},
\bauthor{\bsnm{Mours}, \binits{M.}}:
\batitle{Rheology of polymers near liquid-solid transitions}.
\bjtitle{Adv. Polym. Sci.}
\bvolume{134},
\bfpage{167}--\blpage{230}
(\byear{1997})
\end{barticle}
\endbibitem

\bibitem{Matsumoto:1988}
\begin{barticle}
\bauthor{\bsnm{Matsumoto}, \binits{T.}},
\bauthor{\bsnm{Mashiko}, \binits{K.}}:
\batitle{Influence of added salt on dynamic viscoelasticity of
  carboxymethylcellulose aqueous systems}.
\bjtitle{Polym. Eng. Sci.}
\bvolume{28}(\bissue{6}),
\bfpage{393}--\blpage{402}
(\byear{1988})
\end{barticle}
\endbibitem

\bibitem{Kastner:1997}
\begin{barticle}
\bauthor{\bsnm{K{\"a}stner}, \binits{U.}},
\bauthor{\bsnm{Hoffmann}, \binits{H.}},
\bauthor{\bsnm{D{\"o}nges}, \binits{R.}},
\bauthor{\bsnm{Hilbig}, \binits{J.}}, \betal:
\batitle{Structure and solution properties of sodium carboxymethyl cellulose}.
\bjtitle{Colloids Surf.}
\bvolume{123},
\bfpage{307}--\blpage{328}
(\byear{1997})
\end{barticle}
\endbibitem

\bibitem{Benchabane:2008}
\begin{barticle}
\bauthor{\bsnm{Benchabane}, \binits{A.}},
\bauthor{\bsnm{Bekkour}, \binits{K.}}:
\batitle{Rheological properties of carboxymethyl cellulose ({CMC}) solutions}.
\bjtitle{Colloid Polym. Sci.}
\bvolume{286}(\bissue{10}),
\bfpage{1173}--\blpage{1180}
(\byear{2008})
\end{barticle}
\endbibitem

\bibitem{Lopez:2019}
\begin{barticle}
\bauthor{\bsnm{Lopez}, \binits{C.G.}},
\bauthor{\bsnm{Richtering}, \binits{W.}}:
\batitle{Influence of divalent counterions on the solution rheology and
  supramolecular aggregation of carboxymethyl cellulose}.
\bjtitle{Cellulose}
\bvolume{26},
\bfpage{1517}--\blpage{1534}
(\byear{2019})
\end{barticle}
\endbibitem

\bibitem{Jaishankar:2013}
\begin{barticle}
\bauthor{\bsnm{Jaishankar}, \binits{A.}},
\bauthor{\bsnm{McKinley}, \binits{G.H.}}:
\batitle{Power-law rheology in the bulk and at the interface: quasi-properties
  and fractional constitutive equations}.
\bjtitle{Proc. R. Soc. A: Math. Phys. Eng. Sci.}
\bvolume{469}(\bissue{2149}),
\bfpage{20120284}
(\byear{2013})
\end{barticle}
\endbibitem

\bibitem{Bonfanti:2020}
\begin{barticle}
\bauthor{\bsnm{Bonfanti}, \binits{A.}},
\bauthor{\bsnm{Kaplan}, \binits{J.L.}},
\bauthor{\bsnm{Charras}, \binits{G.}},
\bauthor{\bsnm{Kabla}, \binits{A.}}:
\batitle{Fractional viscoelastic models for power-law materials}.
\bjtitle{Soft Matter}
\bvolume{16}(\bissue{26}),
\bfpage{6002}--\blpage{6020}
(\byear{2020})
\end{barticle}
\endbibitem

\bibitem{Larson:1999}
\begin{bbook}
\bauthor{\bsnm{Larson}, \binits{R.G.}}:
\bbtitle{The Structure and Rheology of Complex Fluids}
vol. \bseriesno{150}.
\bpublisher{Oxford university press New York}, \blocation{???}
(\byear{1999})
\end{bbook}
\endbibitem

\bibitem{leibler1991dynamics}
\begin{barticle}
\bauthor{\bsnm{Leibler}, \binits{L.}},
\bauthor{\bsnm{Rubinstein}, \binits{M.}},
\bauthor{\bsnm{Colby}, \binits{R.H.}}:
\batitle{Dynamics of reversible networks}.
\bjtitle{Macromolecules}
\bvolume{24}(\bissue{16}),
\bfpage{4701}--\blpage{4707}
(\byear{1991})
\end{barticle}
\endbibitem

\bibitem{rubinstein2001dynamics}
\begin{barticle}
\bauthor{\bsnm{Rubinstein}, \binits{M.}},
\bauthor{\bsnm{Semenov}, \binits{A.N.}}:
\batitle{Dynamics of entangled solutions of associating polymers}.
\bjtitle{Macromolecules}
\bvolume{34}(\bissue{4}),
\bfpage{1058}--\blpage{1068}
(\byear{2001})
\end{barticle}
\endbibitem

\bibitem{rubinstein2003polymer}
\begin{bbook}
\bauthor{\bsnm{Rubinstein}, \binits{M.}},
\bauthor{\bsnm{Colby}, \binits{R.H.}}:
\bbtitle{Polymer Physics}.
\bpublisher{Oxford university press}, \blocation{???}
(\byear{2003})
\end{bbook}
\endbibitem

\bibitem{Matsumoto:1990}
\begin{barticle}
\bauthor{\bsnm{Matsumoto}, \binits{T.}},
\bauthor{\bsnm{Ito}, \binits{D.}}:
\batitle{Viscoelastic and nuclear magnetic resonance studies on molecular
  mobility of carboxymethylcellulose--calcium complex in concentrated aqueous
  systems}.
\bjtitle{J. Chem. Soc., Faraday Trans.}
\bvolume{86}(\bissue{5}),
\bfpage{829}--\blpage{832}
(\byear{1990})
\end{barticle}
\endbibitem

\bibitem{Matsumoto:1992}
\begin{barticle}
\bauthor{\bsnm{Matsumoto}, \binits{T.}},
\bauthor{\bsnm{Zenkoh}, \binits{H.}}:
\batitle{A new molecular model for complexation between carboxymethylcellulose
  and alkaline—earth metal ions in aqueous systems}.
\bjtitle{Food hydrocoll.}
\bvolume{6}(\bissue{4}),
\bfpage{379}--\blpage{386}
(\byear{1992})
\end{barticle}
\endbibitem

\bibitem{stadler2009linear}
\begin{barticle}
\bauthor{\bsnm{Stadler}, \binits{F.J.}},
\bauthor{\bsnm{Pyckhout-Hintzen}, \binits{W.}},
\bauthor{\bsnm{Schumers}, \binits{J.-M.}},
\bauthor{\bsnm{Fustin}, \binits{C.-A.}},
\bauthor{\bsnm{Gohy}, \binits{J.-F.}},
\bauthor{\bsnm{Bailly}, \binits{C.}}:
\batitle{Linear viscoelastic rheology of moderately entangled telechelic
  polybutadiene temporary networks}.
\bjtitle{Macromolecules}
\bvolume{42}(\bissue{16}),
\bfpage{6181}--\blpage{6192}
(\byear{2009})
\end{barticle}
\endbibitem

\bibitem{coutouly2024low}
\begin{botherref}
\oauthor{\bsnm{Coutouly}, \binits{C.}},
\oauthor{\bsnm{Mortensen}, \binits{K.}},
\oauthor{\bparticle{van} \bsnm{Ruymbeke}, \binits{E.}},
\oauthor{\bsnm{Fustin}, \binits{C.-A.}}:
Low t g, strongly segregated, aba triblock copolymers: a rheological and
  structural study.
Soft Matter
(2024)
\end{botherref}
\endbibitem

\bibitem{lyons2022equilibration}
\begin{barticle}
\bauthor{\bsnm{Lyons}, \binits{R.}},
\bauthor{\bsnm{Hammer}, \binits{L.}},
\bauthor{\bsnm{Andr{\'e}}, \binits{A.}},
\bauthor{\bsnm{Fustin}, \binits{C.-A.}},
\bauthor{\bsnm{Nicola{\"y}}, \binits{R.}},
\bauthor{\bsnm{Van~Ruymbeke}, \binits{E.}}:
\batitle{Equilibration dynamics of a dynamic covalent network diluted in a
  metallosupramolecular polymer matrix}.
\bjtitle{J. Rheol.}
\bvolume{66}(\bissue{6}),
\bfpage{1349}--\blpage{1364}
(\byear{2022})
\end{barticle}
\endbibitem

\bibitem{Xu:2019}
\begin{barticle}
\bauthor{\bsnm{Xu}, \binits{Z.}},
\bauthor{\bsnm{Song}, \binits{Y.}},
\bauthor{\bsnm{Zheng}, \binits{Q.}}:
\batitle{Payne effect of carbon black filled natural rubber compounds and their
  carbon black gels}.
\bjtitle{Polymer}
\bvolume{185},
\bfpage{121953}
(\byear{2019})
\end{barticle}
\endbibitem

\bibitem{Fan:2019}
\begin{barticle}
\bauthor{\bsnm{Fan}, \binits{X.}},
\bauthor{\bsnm{Xu}, \binits{H.}},
\bauthor{\bsnm{Zhang}, \binits{Q.}},
\bauthor{\bsnm{Song}, \binits{Y.}},
\bauthor{\bsnm{Zheng}, \binits{Q.}}, \betal:
\batitle{Insight into the weak strain overshoot of carbon black filled natural
  rubber}.
\bjtitle{Polymer}
\bvolume{167},
\bfpage{109}--\blpage{117}
(\byear{2019})
\end{barticle}
\endbibitem

\bibitem{Hyun:2002}
\begin{barticle}
\bauthor{\bsnm{Hyun}, \binits{K.}},
\bauthor{\bsnm{Kim}, \binits{S.H.}},
\bauthor{\bsnm{Ahn}, \binits{K.H.}},
\bauthor{\bsnm{Lee}, \binits{S.J.}}:
\batitle{Large amplitude oscillatory shear as a way to classify the complex
  fluids}.
\bjtitle{J. Non-Newton. Fluid Mech.}
\bvolume{107}(\bissue{1-3}),
\bfpage{51}--\blpage{65}
(\byear{2002})
\end{barticle}
\endbibitem

\bibitem{Miyazaki:2006}
\begin{barticle}
\bauthor{\bsnm{Miyazaki}, \binits{K.}},
\bauthor{\bsnm{Wyss}, \binits{H.M.}},
\bauthor{\bsnm{Weitz}, \binits{D.A.}},
\bauthor{\bsnm{Reichman}, \binits{D.R.}}:
\batitle{Nonlinear viscoelasticity of metastable complex fluids}.
\bjtitle{Europhys. Lett.}
\bvolume{75}(\bissue{6}),
\bfpage{915}
(\byear{2006})
\end{barticle}
\endbibitem

\bibitem{Sudreau:2022}
\begin{barticle}
\bauthor{\bsnm{Sudreau}, \binits{I.}},
\bauthor{\bsnm{Manneville}, \binits{S.}},
\bauthor{\bsnm{Servel}, \binits{M.}},
\bauthor{\bsnm{Divoux}, \binits{T.}}:
\batitle{Shear-induced memory effects in boehmite gels}.
\bjtitle{J. Rheol.}
\bvolume{66}(\bissue{1}),
\bfpage{91}--\blpage{104}
(\byear{2022})
\end{barticle}
\endbibitem

\bibitem{Helgeson:2007}
\begin{barticle}
\bauthor{\bsnm{Helgeson}, \binits{M.E.}},
\bauthor{\bsnm{Wagner}, \binits{N.J.}},
\bauthor{\bsnm{Vlassopoulos}, \binits{D.}}:
\batitle{Viscoelasticity and shear melting of colloidal star polymer glasses}.
\bjtitle{J. Rheol.}
\bvolume{51}(\bissue{2}),
\bfpage{297}--\blpage{316}
(\byear{2007})
\end{barticle}
\endbibitem

\bibitem{Carrier:2009}
\begin{barticle}
\bauthor{\bsnm{Carrier}, \binits{V.}},
\bauthor{\bsnm{Petekidis}, \binits{G.}}:
\batitle{Nonlinear rheology of colloidal glasses of soft thermosensitive
  microgel particles}.
\bjtitle{J. Rheol.}
\bvolume{53}(\bissue{2}),
\bfpage{245}--\blpage{273}
(\byear{2009})
\end{barticle}
\endbibitem

\bibitem{Koumakis:2012}
\begin{barticle}
\bauthor{\bsnm{Koumakis}, \binits{N.}},
\bauthor{\bsnm{Pamvouxoglou}, \binits{A.}},
\bauthor{\bsnm{Poulos}, \binits{A.}},
\bauthor{\bsnm{Petekidis}, \binits{G.}}:
\batitle{Direct comparison of the rheology of model hard and soft particle
  glasses}.
\bjtitle{Soft Matter}
\bvolume{8}(\bissue{15}),
\bfpage{4271}--\blpage{4284}
(\byear{2012})
\end{barticle}
\endbibitem

\bibitem{Cho:2005}
\begin{barticle}
\bauthor{\bsnm{Cho}, \binits{K.S.}},
\bauthor{\bsnm{Hyun}, \binits{K.}},
\bauthor{\bsnm{Ahn}, \binits{K.H.}},
\bauthor{\bsnm{Lee}, \binits{S.J.}}:
\batitle{A geometrical interpretation of large amplitude oscillatory shear
  response}.
\bjtitle{J. Rheol.}
\bvolume{49}(\bissue{3}),
\bfpage{747}--\blpage{758}
(\byear{2005})
\end{barticle}
\endbibitem

\bibitem{Ewoldt:2008}
\begin{barticle}
\bauthor{\bsnm{Ewoldt}, \binits{R.H.}},
\bauthor{\bsnm{Hosoi}, \binits{A.}},
\bauthor{\bsnm{McKinley}, \binits{G.H.}}:
\batitle{New measures for characterizing nonlinear viscoelasticity in large
  amplitude oscillatory shear}.
\bjtitle{J. Rheol.}
\bvolume{52}(\bissue{6}),
\bfpage{1427}--\blpage{1458}
(\byear{2008})
\end{barticle}
\endbibitem

\bibitem{doi1979dynamics}
\begin{barticle}
\bauthor{\bsnm{Doi}, \binits{M.}},
\bauthor{\bsnm{Edwards}, \binits{S.}}:
\batitle{Dynamics of concentrated polymer systems. part 4.—rheological
  properties}.
\bjtitle{J. Chem. Soc., Faraday Trans. 2}
\bvolume{75},
\bfpage{38}--\blpage{54}
(\byear{1979})
\end{barticle}
\endbibitem

\bibitem{nebouy2021flow}
\begin{barticle}
\bauthor{\bsnm{N{\'e}bouy}, \binits{M.}},
\bauthor{\bsnm{Chazeau}, \binits{L.}},
\bauthor{\bsnm{Morthomas}, \binits{J.}},
\bauthor{\bsnm{Fusco}, \binits{C.}},
\bauthor{\bsnm{Dieudonn{\'e}-George}, \binits{P.}},
\bauthor{\bsnm{Baeza}, \binits{G.P.}}:
\batitle{Flow-induced crystallization of a multiblock copolymer under large
  amplitude oscillatory shear: Experiments and modeling}.
\bjtitle{J. Rheol.}
\bvolume{65}(\bissue{3}),
\bfpage{405}--\blpage{418}
(\byear{2021})
\end{barticle}
\endbibitem

\bibitem{Pham:2006}
\begin{barticle}
\bauthor{\bsnm{Pham}, \binits{K.}},
\bauthor{\bsnm{Petekidis}, \binits{G.}},
\bauthor{\bsnm{Vlassopoulos}, \binits{D.}},
\bauthor{\bsnm{Egelhaaf}, \binits{S.}},
\bauthor{\bsnm{Pusey}, \binits{P.}},
\bauthor{\bsnm{Poon}, \binits{W.}}:
\batitle{Yielding of colloidal glasses}.
\bjtitle{Europhys. Lett.}
\bvolume{75}(\bissue{4}),
\bfpage{624}
(\byear{2006})
\end{barticle}
\endbibitem

\bibitem{Donley:2020}
\begin{barticle}
\bauthor{\bsnm{Donley}, \binits{G.J.}},
\bauthor{\bsnm{Singh}, \binits{P.K.}},
\bauthor{\bsnm{Shetty}, \binits{A.}},
\bauthor{\bsnm{Rogers}, \binits{S.A.}}:
\batitle{Elucidating the $g''$ overshoot in soft materials with a yield
  transition via a time-resolved experimental strain decomposition}.
\bjtitle{Proc. Natl. Acad. Sci. U.S.A.}
\bvolume{117}(\bissue{36}),
\bfpage{21945}--\blpage{21952}
(\byear{2020})
\end{barticle}
\endbibitem

\bibitem{Manneville:2008}
\begin{barticle}
\bauthor{\bsnm{Manneville}, \binits{S.}}:
\batitle{Recent experimental probes of shear banding}.
\bjtitle{Rheol. Acta}
\bvolume{47}(\bissue{3}),
\bfpage{301}--\blpage{318}
(\byear{2008})
\end{barticle}
\endbibitem

\bibitem{SaintMichel2016}
\begin{barticle}
\bauthor{\bsnm{Saint-Michel}, \binits{B.}},
\bauthor{\bsnm{Gibaud}, \binits{T.}},
\bauthor{\bsnm{Leocmach}, \binits{M.}},
\bauthor{\bsnm{Manneville}, \binits{S.}}:
\batitle{Local oscillatory rheology from echography}.
\bjtitle{Phys. Rev. Appl.}
\bvolume{5},
\bfpage{034014}
(\byear{2016})
\end{barticle}
\endbibitem

\bibitem{MorletDecarnin:2023}
\begin{barticle}
\bauthor{\bsnm{Morlet-Decarnin}, \binits{L.}},
\bauthor{\bsnm{Divoux}, \binits{T.}},
\bauthor{\bsnm{Manneville}, \binits{S.}}:
\batitle{Critical-like gelation dynamics in cellulose nanocrystal suspensions}.
\bjtitle{ACS Macro Lett.}
\bvolume{12}(\bissue{12}),
\bfpage{1733}--\blpage{1738}
(\byear{2023})
\end{barticle}
\endbibitem

\bibitem{Benzi:2023}
\begin{barticle}
\bauthor{\bsnm{Benzi}, \binits{R.}},
\bauthor{\bsnm{Divoux}, \binits{T.}},
\bauthor{\bsnm{Barentin}, \binits{C.}},
\bauthor{\bsnm{Manneville}, \binits{S.}},
\bauthor{\bsnm{Sbragaglia}, \binits{M.}},
\bauthor{\bsnm{Toschi}, \binits{F.}}:
\batitle{Continuum modeling of soft glassy materials under shear}.
\bjtitle{Europhys. Lett.}
\bvolume{141}(\bissue{5}),
\bfpage{56001}
(\byear{2023})
\end{barticle}
\endbibitem

\bibitem{Kamani:2021}
\begin{barticle}
\bauthor{\bsnm{Kamani}, \binits{K.}},
\bauthor{\bsnm{Donley}, \binits{G.J.}},
\bauthor{\bsnm{Rogers}, \binits{S.A.}}:
\batitle{Unification of the rheological physics of yield stress fluids}.
\bjtitle{Phys. Rev. Lett.}
\bvolume{126}(\bissue{21}),
\bfpage{218002}
(\byear{2021})
\end{barticle}
\endbibitem

\bibitem{Kamani:2024}
\begin{barticle}
\bauthor{\bsnm{Kamani}, \binits{K.M.}},
\bauthor{\bsnm{Rogers}, \binits{S.A.}}:
\batitle{Brittle and ductile yielding in soft materials}.
\bjtitle{Proc. Natl. Acad. Sci. U.S.A.}
\bvolume{121}(\bissue{22}),
\bfpage{2401409121}
(\byear{2024})
\end{barticle}
\endbibitem

\end{thebibliography}
\end{document}